\documentclass[12pt]{article}
\usepackage[a4paper,margin=1in]{geometry}
\usepackage[titletoc,title]{appendix}
\usepackage{changepage}
\usepackage[utf8x]{inputenc}
\usepackage{textcomp,marvosym}
\usepackage{fixltx2e}
\usepackage{amsmath,amssymb}
\usepackage{cite}
\usepackage{nameref,hyperref}
\usepackage{authblk}
\usepackage{algorithm}
\usepackage{algpseudocode}
\usepackage{dsfont}
\usepackage{xcolor}
\usepackage{enumitem}
\usepackage{caption,setspace}
\usepackage{subcaption}
\usepackage{pdflscape}
\usepackage{graphicx}
\usepackage{placeins}
\usepackage{mathtools}

\renewcommand{\eqref}[1]{Equation~(\ref{#1})}

\title{A sharp-front moving boundary model for malignant invasion}

\author[1]{Maud El-Hachem}
\author[1]{Scott~W. McCue}
\author[1]{Matthew~J. Simpson\footnote{To whom correspondence should be addressed. E-mail: matthew.simpson@qut.edu.au}}

\affil[1]{School of Mathematical Sciences, Queensland University of Technology, Brisbane, Queensland 4001, Australia}

\begin{document}

\maketitle
\begin{abstract}
We analyse a novel mathematical model of malignant invasion which takes the form of a two-phase moving boundary problem describing the invasion of a population of malignant cells into a population of background tissue, such as skin.  Cells in both populations undergo diffusive migration and logistic proliferation.  The interface between the two populations moves according to a two-phase Stefan condition.  Unlike many reaction-diffusion models of malignant invasion,  the moving boundary model explicitly describes the motion of the sharp front between the cancer and surrounding tissues without needing to introduce degenerate nonlinear diffusion.  Numerical simulations suggest the model gives rise to very interesting travelling wave solutions that move with speed $c$, and the model supports both malignant invasion and malignant retreat, where the travelling wave can move in either the positive or negative $x$-directions.  Unlike the well-studied Fisher-Kolmogorov and Porous-Fisher models where travelling waves move with a minimum wave speed $c \ge c^* > 0$, the moving boundary model leads to travelling wave solutions with $|c| <  c^{**}$.  We interpret these travelling wave solutions in the phase plane and show that they are associated with several features of the classical Fisher-Kolmogorov phase plane that are often disregarded as being nonphysical.   We show, numerically, that the phase plane analysis compares well with long time solutions from the full partial differential equation model as well as providing  accurate perturbation approximations for the shape of the travelling waves.
\end{abstract}


\newpage
\section{Introduction}
\label{sec:intro}

\begin{figure}[h]
	\centering
	\includegraphics[width=1\textwidth]{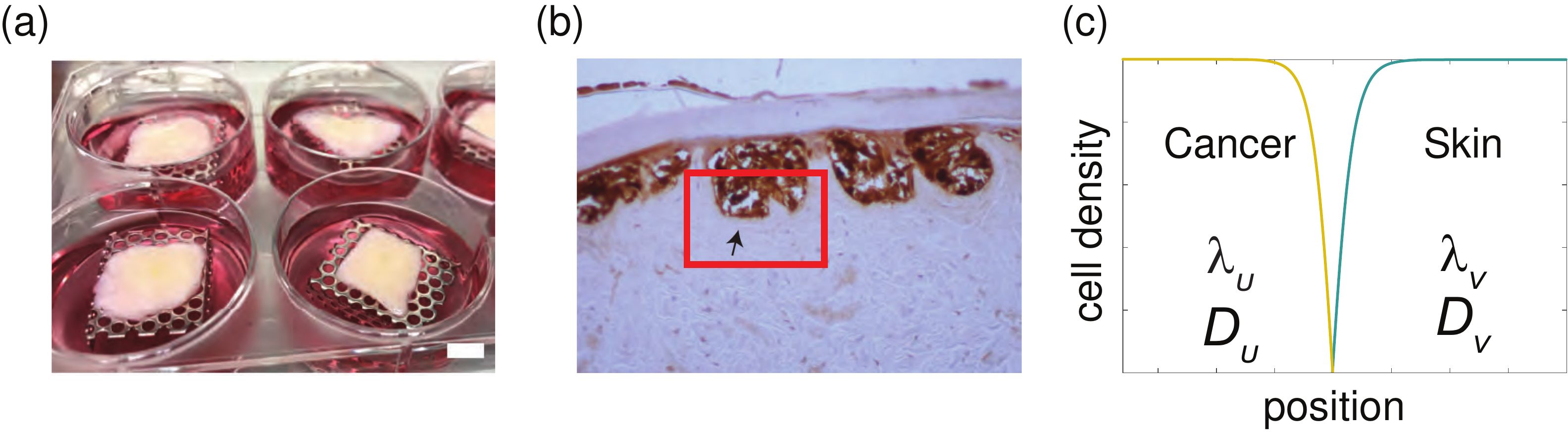}
	\caption{\textbf{Experimental motivation and model schematic}. (a) Experimental protocol where a population of motile and proliferative melanoma cells are placed onto the surface of human skin tissues kept at an air-liquid interface to simulate the \textit{in vivo} environment. Scale bar is 6 mm. (b) Vertical cross section through the tissues in (a) highlighting the vertical downward invasion of melanoma cells (dark) into surrounding skin tissue (light).  The sharp front separating the invading malignant population from the surrounding tissues is visually distinct and highlighted in the red rectangle.  Images in (a)-(b) are reproduced from Haridas~\cite{Haridas2017} with permission. (b) Schematic solution of a one-dimensional partial differential equation solution showing the spatial distribution of a population of cancer cells and skin cells separated by a sharp front.  The cancer cells have density $u(x,t)$, diffusivity $D_u$ and proliferation rate $\lambda_u$.  The skin cells have density $v(x,t)$, diffusivity $D_v$ and proliferation rate $\lambda_v$.}
	\label{fig:figure1}
\end{figure}

Populations of motile and proliferative cells can give rise to moving fronts that are associated with cancer progression and malignant invasion~\cite{Swanson2003,Gatenby1996,Roose2007,Byrne2010}. Similar invasive phenomena are associated with wound healing~\cite{Maini2004a,Maini2004b,Simpson2013}, development~\cite{Simpson2007,Sengers2007} and ecology~\cite{Skellam1951,Shigesada1995,Broadbridge2002,BradshawHajek2004}. Mathematically, these fronts are often studied using reaction-diffusion equations that are based upon the well-known Fisher-Kolmogorov model or extensions~\cite{Fisher1937,Kolmogorov1937,Canosa1973,Sherratt90,Murray02}.  While such models are able to capture certain important features, such as the formation of constant speed travelling wave solutions, there are other features of the standard Fisher-Kolmogorov model that are inconsistent with biological observations.  For example, classical travelling wave solutions of the Fisher-Kolmogorov model on $-\infty < x < \infty$  do not involve a well-defined front because the travelling wave solutions do not have compact support and the cell density is always positive, with $u(x,t) \to 0$ as $x \to  \infty$.  Solutions of the Fisher-Kolmogorov model on $-\infty < x < \infty$ always lead to travelling waves for initial conditions with compact support, and these travelling waves lead to the colonisation of initially-vacant regions without ever retreating.  These two features are inconsistent with many experimental observations.  Experimental images in Figure \ref{fig:figure1}(a)--(b) show key features of malignant invasion.  Here a population of motile and proliferative melanoma cells is placed onto the surface of human skin tissues in Figure \ref{fig:figure1}(a).  A vertical section through the skin tissues show the melanoma invading vertically downward into the surrounding skin cells and we see a clear sharp front between the two subpopulations~\cite{Haridas2017,Haridas2018}. In reality, such fronts can either invade into, or retreat from, the surrounding tissues~\cite{Hanahan2000}.  Neither of these biological features are consistent with travelling wave solutions of the classical Fisher-Kolmogorov model.

One way to extend the Fisher-Kolmogorov model to produce a well-defined front is to introduce nonlinear degenerate diffusion~\cite{Sengers2007,Murray02,Sherratt90,Sanchez1994,Sanchez1995,Witelski1994,Witelski1995,Sherratt1996,Harris2004,Jin2016,Warne2019,McCue2019}.  Such models, often called the Porous-Fisher model, give rise to travelling wave solutions with a well-defined sharp front that always lead to advancing travelling waves that never retreat.  One potential weakness of this approach is that the introduction of nonlinear degenerate diffusion leads to additional model parameters that can be difficult to estimate and interpret~\cite{Sherratt90,Warne2019,McCue2019,Simpson2011}. Another way to introduce a sharp front into the Fisher--Kolmogorov model is to recast the problem as a moving boundary problem~\cite{Crank1987,Hill1987,Gupta2017,McCue2008}.  This approach involves studying the Fisher-Kolmogorov model on  $0 < x < L(t)$, and specifying that $u(L(t),t)=0$ to give a well-defined front.  In this approach a Stefan-condition is applied to determine the speed of the moving front~\cite{Crank1987,Hill1987,Gupta2017,McCue2008}.  Such models, sometimes called the Fisher-Stefan model~\cite{Elhachem2019,Simpson2020,Fadai2020}, have been extensively studied using rigorous analysis~\cite{Du2010,Du2011,Bunting2012,Du2012,Du2014a,Du2014b,Du2015} but have received far less attention in terms of how the solutions of such free boundary problems relate to biological observations.  Interestingly, while free boundary problems are routinely used to study many  problems in industrial and applied mathematics~\cite{Font2013,Mitchell2014,Dalwadi2020}, they are less frequently encountered in the mathematical biology literature.

Of course, a key difference between the classical Fisher-Kolmogorov model and the kinds of applications in Figure \ref{fig:figure1}(a)-(b) is that the usual Fisher-Kolmogorov model deals with just one population of cells, whereas malignant invasion involves one population of cells invading into another population of cells.  To model such applications, the Fisher-Kolmogorov model can be extended to a system of partial differential equations to represent the different cell types present~\cite{Gatenby1996,Landman1998,Perumpanani1999,Painter2003,Simpson2006,Browning2019}.  While the Fisher-Kolmogorov and Porous-Fisher models have been extended to deal with multiple interacting populations, the underlying issues associated with the single population models, described above, also apply to the multiple population analogue~\cite{Simpson2006}.

In this work we study a mathematical model of cell invasion that involves describing two populations of cells as a moving boundary problem.  A schematic of this model in Figure \ref{fig:figure1}(c) shows that we consider two cell populations, such as a population of cancer cells invading into a population of skin cells, which is consistent with the experimental images in Figure \ref{fig:figure1}(a)-(b).  Cells in both populations undergo linear diffusion and proliferate logistically.  The motion of the sharp front is governed by a two-phase Stefan condition~\cite{Crank1987,Hill1987,Gupta2017,McCue2008,Mitchell2014b,Mitchell2015,Mitchell2016}. As we will show, various properties of the solutions of this model are consistent with experimental observations.  Namely, this model leads to a well-defined front and travelling wave solutions that represent either malignant advance or retreat.  It is interesting that the travelling wave analysis of this model is intimately related with the classical phase plane associated with travelling wave solutions of the Fisher-Kolmogorov model.  However, for our model we make use of certain trajectories in the classical phase plane that are normally discarded on the grounds of being nonphysical.  Here, in the context of a moving boundary problem, these normally-discarded features play key roles in determining the travelling wave solutions.

\section{Results and Discussion} \label{sec:ResultsDiscussion}
From this point forward all dimensional variables and parameters are denoted with a circumflex, whereas nondimensional quantities are denoted using regular symbols.

\subsection{Mathematical model} \label{sec:MathModel}
We consider a reaction-diffusion model of a population of cancer cells with density $\hat{u}(\hat{x},\hat{t})$, and a population of skin cells with density $\hat{v}(\hat{x},\hat{t})$.  The system of equations can be written as
\begin{align}
\label{eq:PartialDiffU}
&\frac{\partial \hat{u}}{\partial \hat{t}} =\hat{D}_u \frac{\partial^{2} \hat{u}}{\partial \hat{x}^{2}} + \hat{\lambda}_u \hat{u} \left(1-\dfrac{\hat{u}}{\hat{K}_u}\right), \quad -\hat{L}_u < \hat{x} < \hat{s}(\hat{t}), \\
\label{eq:PartialDiffV}
&\frac{\partial \hat{v}}{\partial \hat{t}} =\hat{D}_v \frac{\partial^{2} \hat{v}}{\partial \hat{x}^{2}}   + \hat{\lambda}_v \hat{v} \left(1-\dfrac{\hat{v}}{\hat{K}_v}\right), \quad  \hat{s}(\hat{t}) < \hat{x} < \hat{L}_v,
\end{align}
where the densities are functions of position, $\hat{x}$, and time, $\hat{t}$. Cancer cells undergo diffusive migration with diffusivity $\hat{D}_u > 0$, and proliferate logistically with rate $\hat{\lambda}_u > 0$ and carrying capacity density $\hat{K}_u > 0$.  Similarly, skin cells undergo diffusive migration with diffusivity $\hat{D}_v > 0$ and proliferate logistically with rate $\hat{\lambda}_v > 0$ and carrying capacity density $\hat{K}_v > 0$.   The model is defined on the $\hat{L}_u < \hat{x} < \hat{L}_v$, with a moving boundary $\hat{x} = \hat{s}(\hat{t})$ separating the population of cancer cells, $\hat{x} < \hat{s}(\hat{t})$, from the population of skin cells, $\hat{x} > \hat{s}(\hat{t})$.

Since we are interested in cell invasion we focus on travelling wave solutions of Equations (\ref{eq:PartialDiffU})-(\ref{eq:PartialDiffV}) by setting $\hat{L}_u$ and $\hat{L}_v$ to be sufficiently large to model an infinite domain problem.  The boundary conditions we consider are
\begin{align} \label{eq:BCNeumann}
&\left.\frac{\partial \hat{u}}{\partial \hat{x}}\right|_{\hat{x}=-\hat{L}_u} = 0, \qquad \qquad \left.\frac{\partial \hat{v}}{\partial \hat{x}}\right|_{\hat{x}=\hat{L}_v} = 0,\\
&\hat{u}(\hat{s}(\hat{t}), \hat{t}) = 0, \qquad \qquad\hat{v}(\hat{s}(\hat{t}), \hat{t}) = 0. \label{eq:BCDirichlet}
\end{align}
This means that we have no flux of cancer cells at the left-most boundary and no flux of skin cells at the right-most boundary, and the density of both populations is zero at the moving boundary, as in Figure \ref{fig:figure1}(a).

We describe the motion of the moving boundary by a two-phase Stefan condition,
\begin{equation}
\frac{\mathrm{d} \hat{s}(\hat{t})}{\mathrm{d} \hat{t}} = -\hat{\kappa_{u}}\left.\frac{\partial \hat{u}}{\partial \hat{x}}\right|_{\hat{x}=\hat{s}(\hat{t})}-\hat{\kappa_{v}}\left.\frac{\partial \hat{v}}{\partial \hat{x}}\right|_{\hat{x}=\hat{s}(\hat{t})}.
\label{eq:StefanCondition}
\end{equation}
Here the speed of the moving boundary is the sum of two terms: the first term on the right of Equation (\ref{eq:StefanCondition}) is proportional to the spatial gradient of the cancer cell density at the moving boundary,
$\hat{x} = \hat{s}(\hat{t})$, and the second term on the right of Equation (\ref{eq:StefanCondition}) is proportional to the spatial gradient of the skin cell density at the moving boundary, $\hat{x} = \hat{s}(\hat{t})$.  The constants of proportionality, $\hat{\kappa_{u}}$ and $\hat{\kappa_{u}}$, play an important role in relating the shape of the density profiles to the speed of the interface.  We will consider the relationship between these constants and the speed of the interface later.

In this work we consider initial conditions given by
\begin{align}
\label{eq:ICPhi}
&\hat{u}(\hat{x}, 0)= \hat{\phi}(\hat{x}) \quad \text { on }  \quad -\hat{L}_u < \hat{x} < \hat{s}(\hat{t}), \\
&\hat{v}(\hat{x}, 0)= \hat{\psi}(\hat{x}) \quad \text { on } \quad  \hat{s}(\hat{t}) < \hat{x} <\hat{L}_v,
\label{eq:ICPsi}
\end{align}
such that $\hat{\phi}(\hat{s}(0)) = \hat{\psi}(\hat{s}(0)) = 0$.

\subsection{Nondimensional model}\label{sec:Nondimensionalisation}
We nondimensionalise the dependent variables by writing $u = \hat{u}/\hat{K}_u$ and  $v = \hat{v}/\hat{K}_v$, and we nondimensionalise the independent variables by writing $x = \hat{x} \sqrt{\hat{\lambda}_u / \hat{D}_u}$ and $t = \hat{\lambda}_u \hat{t}$.  In this nondimensional framework our model can be written as
\begin{align}
\label{eq:PartialDiffUNonDim}
&\frac{\partial u}{\partial t} =  \frac{\partial^{2} u}{\partial x^{2}} + u (1-u), \quad -L_u < x < s(t), \\
\label{eq:PartialDiffVNonDim}
&\frac{\partial v}{\partial t} = D \frac{\partial^{2} v}{\partial x^{2}} + \lambda v (1-v), \quad  s(t) < x < L_v,
\end{align}
where the boundary conditions are given by
\begin{align}
\label{eq:BCNeumannNonDim}
&\left. \frac{\partial u}{\partial x} \right|_{x=-L_u} = 0,  \qquad\left. \frac{\partial v}{\partial x} \right|_{x=L_v} = 0, \\
&u(s(t),t) = 0,  \qquad v(s(t),t) = 0, \\
&\frac{\mathrm{d} s(t)}{\mathrm{d} t} = -\kappa_u  \left. \frac{\partial v}{\partial x} \right|_{x=s(t)} - \kappa_v \left. \frac{\partial u}{\partial x} \right|_{x=s(t)}. \label{eq:NondimbcDirichlet}
\end{align}
The nondimensional model has four parameters,
\begin{align}
\label{eq:DimensionlessParam}
D = \frac{\hat{D}_v}{\hat{D}_u }, \quad \lambda = \frac{\hat{\lambda}_v}{\hat{\lambda}_u}, \quad \kappa_u = \frac{\hat{\kappa}_u \hat{K}_u}{\hat{D}_u}, \quad  \kappa_v = \frac{\hat{\kappa}_v \hat{K}_v}{\hat{D}_u}.
\end{align}
In this framework, $D$ is a relative diffusivity, and setting $D=1$ means that the cancer cells and skin cells are equally motile.  In contrast, setting $D>1$ means that skin cells are more motile than cancer cells, while setting $D<1$ models the situation where skin cells are less motile than cancer cells.  Similar interpretations can be made for the relative proliferation rate $\lambda$.

\begin{figure}[h]
	\centering
	\includegraphics[width=1\textwidth]{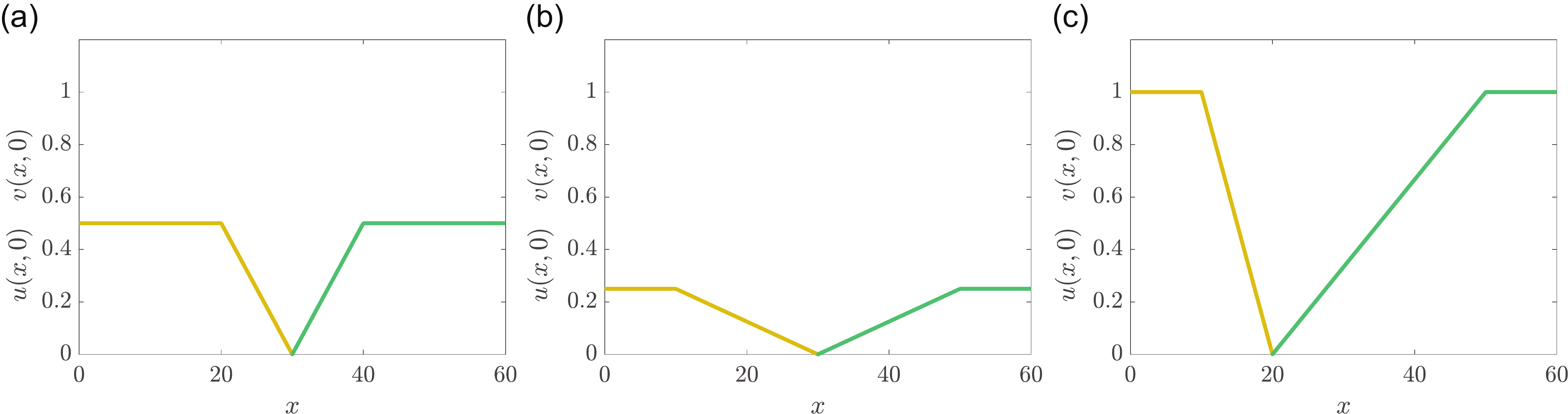}
	\caption{\textbf{Initial condition}.  Three initial conditions on $0 < x < 60$ are shown for: (a) $\alpha = 0.5$, $\beta =  20$ and $s(0) = 30$; (b) $\alpha = 0.25$, $\beta =  10$ and $s(0) = 30$; and (c) $\alpha = 1$, $\beta =  10$ and $s(0) = 20$.}
	\label{fig:figure2}
\end{figure}

We consider numerical solutions of Equations (\ref{eq:PartialDiffUNonDim})-(\ref{eq:PartialDiffVNonDim}) on a domain with $L_u=0$ and $L_v=L$, where $L$ is chosen to be sufficiently large to facilitate the numerical simulation of travelling wave solutions.   We chose piecewise initial conditions given by
\begin{align}
\label{eq:PhiFunctions}
	&u(x,0) = \phi(x) =
	\begin{cases}
		\alpha,  &  0 < x < \beta,\\
		\alpha\left(1-\dfrac{x-\beta}{s(0)-\beta}\right),  &  \beta < x < s(0),
	\end{cases} 	\\
\label{eq:PsiFunctions}
	&v(x,0) = \psi(x) =
	\begin{cases}
	\alpha\left(\dfrac{x-s(0)}{L - \beta - s(0)}\right), & s(0) < x < L-\beta,\\
	\alpha, & L-\beta < x < L,
	\end{cases}
\end{align}
where the parameters $\alpha>0$ and $\beta>0$ control the shape of the piecewise initial density profile.  These initial conditions correspond to  an initial density of $\alpha$ when we are well away from the interface, $x = s(0)$.  Near the interface we set the density to be a linear function of position.  Typical initial conditions in Figure \ref{fig:figure2}  show how varying $\alpha$, $\beta$ and $s(0)$ affects the shape of the initial condition.

\subsection{Numerical solution}\label{sec:Numerical}
To solve Equations (\ref{eq:PartialDiffUNonDim})-(\ref{eq:PartialDiffVNonDim}) we use boundary fixing transformations to recast the moving boundary problem on two fixed domains.  These transformations, $\xi = x/s(t)$ and $\eta = (x-s(t))/(L-s(t))  + 1$, allow us to re-write Equations (\ref{eq:PartialDiffUNonDim})-(\ref{eq:PartialDiffVNonDim}) as,
\begin{align}
\label{eq:PartialDiffUXi}
&\frac{\partial u}{\partial t} = \frac{1}{s^{2}(t)} \frac{\partial^{2} u}{\partial \xi^{2}}+\frac{\xi}{s(t)} \frac{\mathrm{d} s(t)}{\mathrm{d} t} \frac{\partial u}{\partial \xi} + u(1-u), \quad 0 < \xi < 1, \\
&\frac{\partial v}{\partial t} = \frac{D}{\left(L-s(t)\right)^{2}} \frac{\partial^{2} v}{\partial \eta^2}  + \left(\dfrac{2-\eta}{L-s(t)} \right) \frac{\mathrm{d}s(t)}{\mathrm{d} t} \frac{\partial v}{\partial \eta} + \lambda v(1-v), \quad 1 < \eta < 2, \label{eq:PartialDiffVEta}
\end{align}
so that we now have $u(\xi,t)$ on  $ 0 < \xi < 1$ and $v(\eta,t)$ on $1 < \eta < 2$.  The transformed boundary conditions are
\begin{align}
\label{eq:BCNeumannNonDim}
&\left. \frac{\partial u}{\partial  \xi} \right|_{\xi=0} = 0,  \qquad \left. \frac{\partial v}{\partial \eta} \right|_{\eta=2} = 0, \\
&u(1,t) = 0,  \qquad v(1,t) = 0, \\
&\frac{\mathrm{d} s(t)}{\mathrm{d} t} = -\dfrac{\kappa_{u}}{s(t)}\left.\frac{\partial u}{\partial \xi}\right|_{\xi=1} - \dfrac{\kappa_{v}}{L-s(t)}\left.\frac{\partial v }{\partial \eta}\right|_{\eta=1}.
\end{align}
Equations (\ref{eq:PartialDiffUXi})-(\ref{eq:PartialDiffVEta}) and the associated boundary conditions can now be solved numerically using a standard central difference approximation for the transformed spatial derivatives and a backward Euler approximation for the temporal derivatives.  These details are given in Appendix A.

\subsection{Travelling wave solutions}
\begin{landscape}
	\begin{figure}
		\centering
		\includegraphics[width=1.5\textwidth]{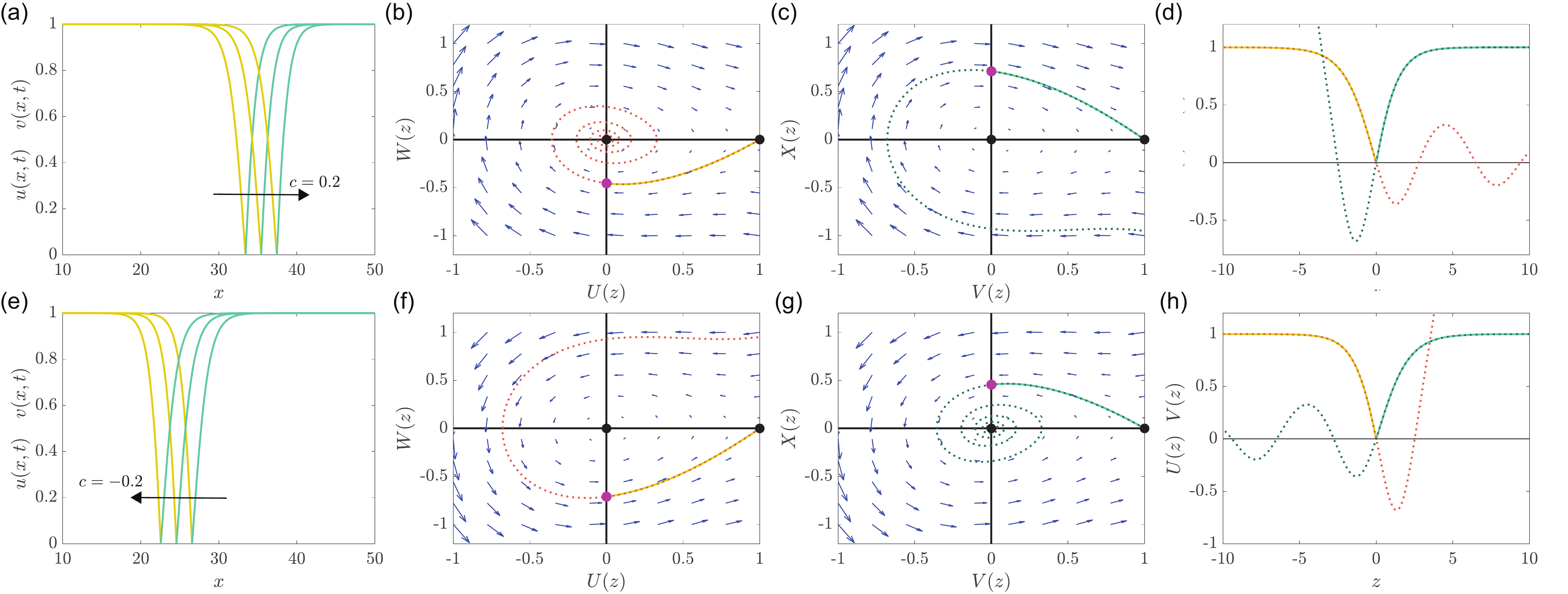}
		\caption{\textbf{Travelling wave solutions for $D = \lambda = 1$}. All partial differential equation solutions are obtained with $L=60$, $\beta=1$, $\alpha = 0.5$, $s(0)=30$.  Results in (a) correspond to $\kappa_u = 1.2195$ and $\kappa_v = 0.5$.  Results in (e) correspond to $\kappa_u = 0.5$ and  $\kappa_v = 1.2195$. Results in (a)-(d) correspond to $c=0.2$ and results in (e)-(h) correspond to $c=-0.2$. Solutions of Equations (\ref{eq:PartialDiffUNonDim})-(\ref{eq:PartialDiffVNonDim}) in (a) and (e) show $u(x,t)$ in solid yellow and $v(x,t)$ in solid green, at $t=20, 30$ and 40.  Phase planes in (b) and (f), and (c) and (g) show the trajectories corresponding to the $U(z)$ and $V(z)$ travelling waves, respectively.  Relevant trajectories in (b)-(c) and (f)-(g) are shown in dashed lines upon which we superimpose the solid lines from the numerical solution of Equations (\ref{eq:PartialDiffUNonDim})-(\ref{eq:PartialDiffVNonDim}) transformed into travelling wave coordinates.  Results in (d) and (h) show $U(z)$ and $V(z)$ as a function of $z$ where results from the phase plane are given in dashed lines superimposed upon the solutions of Equations (\ref{eq:PartialDiffUNonDim})-(\ref{eq:PartialDiffVNonDim}) shifted so that the moving boundary is at $z=0$.  In the phase planes the equilibrium points are shown with a black disc and the intersection of the phase plane trajectory with the vertical axis is shown with a pink disc.}
		\label{fig:figure3}
	\end{figure}
\end{landscape}

Typically, we find that numerical solutions of Equations (\ref{eq:PartialDiffUNonDim})-(\ref{eq:NondimbcDirichlet}) evolve into constant speed, constant shape travelling waves, such as those shown in Figure \ref{fig:figure3}(a).  In this case we have $D=\lambda=1$ so that the cancer cells and skin cells are equally motile and proliferative.  The travelling wave profiles in Figure \ref{fig:figure3}(a) are generated by choosing particular values of $\kappa_u$ and $\kappa_v$ that leads to an invading malignant population moving with positive speed, $c=0.2$.  In contrast, choosing different values of  $\kappa_u$ and $\kappa_v$ can lead to a retreating malignant front, as in Figure \ref{fig:figure3}(e), where we have a travelling wave with $c=-0.2$.  These two numerical travelling wave solutions in Figure \ref{fig:figure3}(a) and (e) are  interesting, especially when we compare the properties of these travelling waves with the more familiar properties of the travelling wave solutions of the  Fisher-Kolmogorov model where there are three important differences:
\begin{enumerate} \itemsep=0mm
\item  The moving boundary model (\ref{eq:PartialDiffUNonDim})-(\ref{eq:NondimbcDirichlet})  supports travelling wave solutions with well-defined sharp front whereas the Fisher-Kolmogorov model does not;\\
\item Travelling wave solutions of the moving boundary model (\ref{eq:PartialDiffUNonDim})-(\ref{eq:NondimbcDirichlet})  can either advance or retreat, whereas analogous travelling wave solutions of the Fisher-Kolmogorov model only ever advance; \\
\item Travelling wave solutions of the nondimensional moving boundary model (\ref{eq:PartialDiffUNonDim})-(\ref{eq:NondimbcDirichlet})  move with speed $|c| < 2$ whereas travelling wave solutions of the nondimensional Fisher-Kolmogorov model always lead to $c \ge  2$.
\end{enumerate}
To provide further insight into the properties of the travelling wave solutions of Equations (\ref{eq:PartialDiffUNonDim})-(\ref{eq:PartialDiffVNonDim}) we now use phase plane analysis.

\subsection{Phase plane analysis}\label{sec:PP}
To study travelling wave solutions of the moving boundary model we re-write the governing equations in the travelling wave coordinate, $z = x-ct$, and seek solutions of the form $U(z)=u(x-ct)$ and $V(z) = v(x-ct)$.  Writing Equations (\ref{eq:PartialDiffUNonDim})-(\ref{eq:PartialDiffVNonDim}) in the travelling wave coordinates leads to
\begin{align} \label{eq:ODEUz}
\frac{\mathrm{d}^2 U}{\mathrm{d} z^2} + c \frac{\mathrm{d} U}{\mathrm{d} z} + U(1-U) &= 0,  \quad  -\infty < z < 0,\\
D \frac{\mathrm{d}^2 V}{\mathrm{d} z^2} + c \frac{\mathrm{d} V}{\mathrm{d} z} + \lambda V(1-V) &= 0,   \quad   0 < z < \infty,
\label{eq:ODEVz}
\end{align}
where the relevant boundary conditions are
\begin{align}
U(-\infty) &= 1, \quad   U(0)= 0, \label{eq:ODEU_BC} \\
V(0) &= 0, \quad  V(\infty) = 1, \label{eq:ODEV_BC} \\
c &=  -\kappa_v \frac{\mathrm{d} V(0)}{\mathrm{d} z}  -\kappa_u \frac{\mathrm{d} U(0)}{\mathrm{d} z}. \label{eq:ODEStefan}
\end{align}
The travelling wave solution for the cancer population, $U(z)$, is described by Equations (\ref{eq:ODEUz}) and (\ref{eq:ODEU_BC}), while the travelling wave solution for the skin population, $V(z)$, is described by Equations (\ref{eq:ODEVz}) and  (\ref{eq:ODEV_BC}).  These two travelling waves are coupled through Equation (\ref{eq:ODEStefan}), which is associated with the Stefan condition at the moving interface.  This means that the travelling wave solutions for $U(z)$ and $V(z)$ can be studied in two separate phase planes, and these two phase planes are coupled by  Equation (\ref{eq:ODEStefan}).

To simplify our study of these two phase planes we note that Equation (\ref{eq:ODEUz}) for $U(z)$ is identical to Equation (\ref{eq:ODEVz}) for $V(z)$ when $D = \lambda = 1$.  Therefore, it is sufficient for us to study Equation (\ref{eq:ODEVz}) for $V(z)$ and to recall that setting $D = \lambda = 1$ means that our analysis of this phase plane corresponds to $U(z)$.  To make progress we re-write Equation (\ref{eq:ODEVz}) as a first-order system
\begin{align} \label{eq:ODESysVXz}
&\dfrac{\mathrm{d} V}{\mathrm{d} z} =  X,  \\
 \label{eq:XVsysdX}
&\dfrac{\mathrm{d} X}{\mathrm{d} z} =  -\dfrac{c}{D}X - \dfrac{\lambda}{D}V(1-V).
\end{align}
At this point we remark that Equation (\ref{eq:ODESysVXz})-(\ref{eq:XVsysdX}) defines a two-dimensional phase plane for $(V(z), X(z))$ that is identical to the phase plane associated with the well-studied travelling wave solutions of the Fisher-Kolmogorov model~\cite{Canosa1973,Murray02}.  Therefore,  all the well-known properties of that phase plane will play a role here in our study of  Equations (\ref{eq:PartialDiffUNonDim})-(\ref{eq:PartialDiffVNonDim}).  In particular, the equilibrium points are $(0,0)$ and $(1,0)$.  Linear stability shows that $(1,0)$ is a saddle for all $c$ whereas $(0,0)$ is a stable node if $c \ge 2\sqrt{\lambda D}$ or a stable spiral for $c < 2\sqrt{\lambda D}$.  Normally, when considering travelling wave solutions of the Fisher-Kolmogorov model, we are interested in the heteroclinic trajectory between $(1,0)$ and $(0,0)$, and the heteroclinic trajectory  associated with the stable spiral at $(0,0)$ when $c < 2\sqrt{\lambda D}$ is ruled out on the basis of requiring $V(z) > 0$.  This classical argument gives rise to the well-known condition that  $c \ge 2\sqrt{\lambda D}$ for travelling wave solutions of the Fisher-Kolmogorov model~\cite{Canosa1973,Murray02}.  In contrast, for our moving boundary model we have a very different situation where, for example, the travelling wave in Figure \ref{fig:figure3}(a) leads to $c = 0.2 < 2 \sqrt{\lambda D}$.

To explore these solutions we show the phase plane corresponding to the travelling wave in Figure \ref{fig:figure3}(a) in Figure \ref{fig:figure3}(b)-(c) for $U(z)$ and $V(z)$ population, respectively. In all phase planes, we generate the trajectories numerically using techniques described in Appendix A.   Figure \ref{fig:figure3}(b) shows the $(U(z), W(z))$ phase plane, where $W(z) = \mathrm{d} U(z) / \mathrm{d} z$ and $c=0.2$ to correspond with the travelling waves in Figure \ref{fig:figure3}(a).  The equilibrium points at $(1,0)$ and $(0,0)$ are highlighted, and the heteroclinic trajectory that leaves $(1,0)$ and spirals into $(0,0)$ is shown with a dotted line.  Normally, when considering travelling wave solutions of the Fisher-Kolmogorov model, this heteroclinic trajectory would be regarded as nonphysical since it implies that $U(z) < 0$ for certain values of $z$ along that trajectory.  However, instead of rejecting this trajectory, the travelling wave solution for $U(z)$ in our moving boundary model simply corresponds to the portion of that heteroclinic trajectory in the fourth quadrant where $U(z) \ge 0$.  The point where the trajectory intersects the $U(z)=0$ axis corresponds to the slope of the travelling wave at the moving boundary,  $(0, W^*(z))$.  This point of intersection is important because it plays a role in satisfying Equation (\ref{eq:ODEStefan}).  To provide an additional check on our phase plane in Figure \ref{fig:figure3}(b) we take the $u(x,t)$ travelling wave profile in Figure \ref{fig:figure3}(a) and superimpose the $(U(z), W(z))$ profile calculated from that travelling wave  as a solid line in the phase plane.  This exercise shows that this solid curve is visually indistinguishable from the first part of the heteroclinic trajectory where $U(z) \ge 0$.

Figure \ref{fig:figure3}(c) shows the $(V(z), X(z))$ phase plane associated with the $v(x,t)$ travelling wave profile in Figure \ref{fig:figure3}(a).  Again, we highlight the equilibrium points at $(1,0)$ and $(0,0)$ and we show the trajectory moving towards the saddle point at $(1,0)$ along the stable manifold.  In the usual study of the Fisher-Kolmogorov model this trajectory is not normally considered because it is not associated with a heteroclinic trajectory, and indeed the phase plane in Figure \ref{fig:figure3}(c) indicates that this trajectory originates far away from the relevant region of the phase plane.  However, we find that part of the trajectory in the first quadrant, just before $(1,0)$ where $V(z) \ge 0$, corresponds to the travelling wave solution for the $v(x,t)$ population.   The point at which this trajectory intersects the $V(z)=0$ axis, $(0, X^*(z))$, corresponds to the slope of the travelling wave at the moving boundary.  Taking the two phase planes in Figure \ref{fig:figure3}(b)-(c) together, the two intersection points $W^*(z)$ and $X^*(z)$ are such that they satisfy Equation (\ref{eq:ODEStefan}), $c = -\kappa_v X^*(z) - \kappa_u W^*(z)$.  Therefore, these two intersection points play a critical role in relating the speed of the travelling wave solution with the constants $\kappa_u$ and $\kappa_v$.

To summarise the results in Figure \ref{fig:figure3}(a)-(c), and to make an explicit connection between the physical solutions of the partial differential equation model and the nonphysical features of the phase plane trajectories, we superimpose various solutions in Figure \ref{fig:figure3}(d).    The solid green and solid yellow lines in Figure \ref{fig:figure3}(d) show long time solutions of Equations (\ref{eq:PartialDiffUNonDim})-(\ref{eq:NondimbcDirichlet}) that are shifted so that the moving boundary is at $z=0$.  The dashed lines in Figure \ref{fig:figure3}(d) shows the $U(z)$ and $V(z)$ associated with the relevant phase plane trajectories from Figure \ref{fig:figure3}(b)-(c), respectively.  In the case of the $U(z)$ trajectory we see that the shape of the trajectory matches the solution from Equations (\ref{eq:PartialDiffUNonDim})-(\ref{eq:PartialDiffVNonDim}) where $z \le 0$ and $U(z) \ge 0$.  The phase plane trajectory of $U(z)$ for $z > 0$ is nonphysical since $U(z)$ oscillates about $U(z)=0$ and this does not correspond to any part of the solution of Equations (\ref{eq:PartialDiffUNonDim})-(\ref{eq:PartialDiffVNonDim}).  In the case of the $V(z)$ profile we see that the shape of the phase plane trajectory matches the solution from Equations (\ref{eq:PartialDiffUNonDim})-(\ref{eq:PartialDiffVNonDim}) where $z \ge 0$ and $V(z) \ge 0$.  The phase plane trajectory of $V(z)$ for $z < 0$ is nonphysical since part of that trajectory involves $V(z) < 0$.

All results in Figure \ref{fig:figure3}(a)-(d) correspond to choices of $\kappa_u$ and $\kappa_v$ that lead to $c=0.2$.  Results in Figure \ref{fig:figure3}(e)-(h) correspond to different choices of $\kappa_u$ and $\kappa_v$ such that the travelling wave leads to a receding front with $c=-0.2$.  Numerical solutions of Equations (\ref{eq:PartialDiffUNonDim})-(\ref{eq:PartialDiffVNonDim}) in Figure \ref{fig:figure3}(e) show the travelling wave solutions and the phase planes in Figure \ref{fig:figure3}(f)-(g) show the phase plane trajectories associated with the  $U(z)$ and $V(z)$ travelling waves.  Again, a summary comparing the physical travelling wave solutions from Equations (\ref{eq:PartialDiffUNonDim})-(\ref{eq:PartialDiffVNonDim}) with the phase plane trajectories is given in Figure \ref{fig:figure3}(h).  This comparison shows that the travelling wave solutions of Equations (\ref{eq:PartialDiffUNonDim})-(\ref{eq:PartialDiffVNonDim}) compare very well with the physical portion of the phase plane trajectories in Figure \ref{fig:figure3}(f)-(g).

\begin{landscape}
	\begin{figure}
		\centering
		\includegraphics[width=1.5\textwidth]{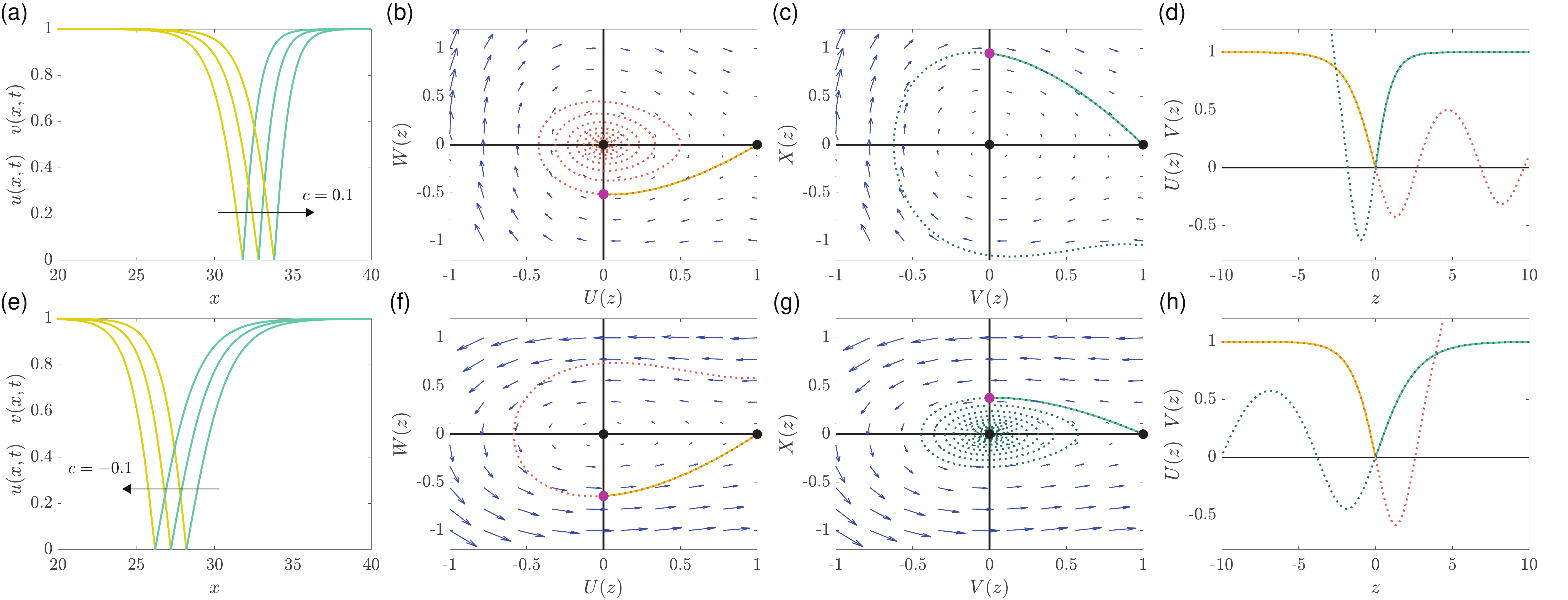}
\caption{\textbf{Travelling wave solutions for $D \ne 1$ and $\lambda = 1$}. All partial differential equation solutions are obtained with $L=60$, $\beta=1$, $\alpha = 0.5$, $s(0)=30$.  Results in (a) correspond to $D=0.5$, $\kappa_u = 1.1151$ and $\kappa_v = 0.5$.  Results in (e) correspond to $D=2$, $\kappa_u = 0.5$ and  $\kappa_v = 1.1180$. Results in (a)-(d) correspond to $c=0.1$ and results in (e)-(h) correspond to $c=-0.1$. Solutions of Equations (\ref{eq:PartialDiffUNonDim})-(\ref{eq:PartialDiffVNonDim}) in (a) and (e) show $u(x,t)$ in solid yellow and $v(x,t)$ in solid green, at $t=20, 30$ and 40.  Phase planes in (b) and (f), and (c) and (g) show the trajectories corresponding to the $U(z)$ and $V(z)$ travelling waves, respectively.  Relevant trajectories in (b)-(c) and (f)-(g) are shown in dashed lines upon which we superimpose the solid lines from the numerical solution of Equations (\ref{eq:PartialDiffUNonDim})-(\ref{eq:PartialDiffVNonDim}) transformed into travelling wave coordinates.  Results in (d) and (h) show $U(z)$ and $V(z)$ as a function of $z$ where results from the phase plane are given in dashed lines superimposed upon the solutions of Equations (\ref{eq:PartialDiffUNonDim})-(\ref{eq:PartialDiffVNonDim}) shifted so that the moving boundary is at $z=0$.  In the phase planes the equilibrium points are shown with a black disc and the intersection of the phase plane trajectory with the vertical axis is shown with a pink disc.}
		\label{fig:figure4}
	\end{figure}
\end{landscape}

\begin{landscape}
\begin{figure}
	\centering
	\includegraphics[width=1.5\textwidth]{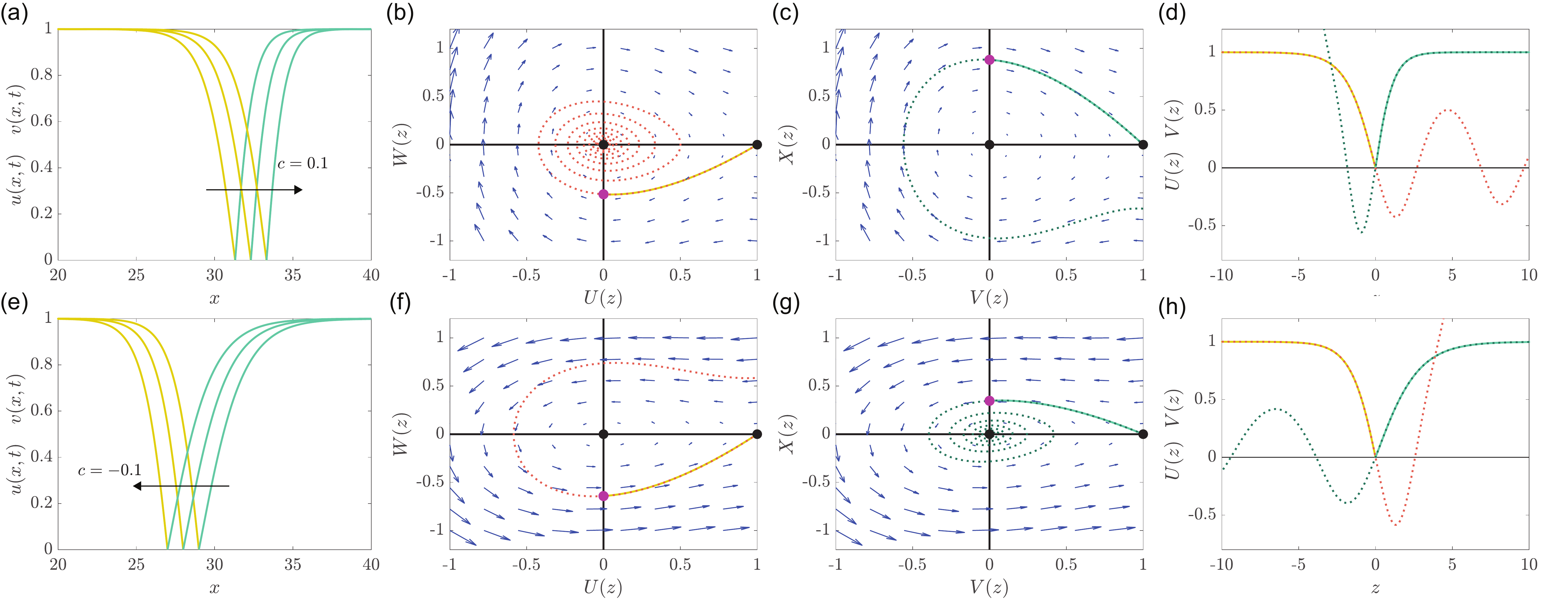}
\caption{\textbf{Travelling wave solutions for $D = 1$ and $\lambda \ne  1$}. All partial differential equation solutions are obtained with $L=60$, $\beta=1$, $\alpha = 0.5$, $s(0)=30$.  Results in (a) correspond to $\lambda=2$, $\kappa_u = 1.0502$ and $\kappa_v = 0.5$.  Results in (e) correspond to $\lambda=0.5$, $\kappa_u = 0.5$ and  $\kappa_v = 1.2164$. Results in (a)-(d) correspond to $c=0.1$ and results in (e)-(h) correspond to $c=-0.1$. Solutions of Equations (\ref{eq:PartialDiffUNonDim})-(\ref{eq:PartialDiffVNonDim}) in (a) and (e) show $u(x,t)$ in solid yellow and $v(x,t)$ in solid green, at $t=20, 30$ and 40.  Phase planes in (b) and (f), and (c) and (g) show the trajectories corresponding to the $U(z)$ and $V(z)$ travelling waves, respectively.  Relevant trajectories in (b)-(c) and (f)-(g) are shown in dashed lines upon which we superimpose the solid lines from the numerical solution of Equations (\ref{eq:PartialDiffUNonDim})-(\ref{eq:PartialDiffVNonDim}) transformed into travelling wave coordinates.  Results in (d) and (h) show $U(z)$ and $V(z)$ as a function of $z$ where results from the phase plane are given in dashed lines superimposed upon the solutions of Equations (\ref{eq:PartialDiffUNonDim})-(\ref{eq:PartialDiffVNonDim}) shifted so that the moving boundary is at $z=0$.  In the phase planes the equilibrium points are shown with a black disc and the intersection of the phase plane trajectory with the vertical axis is shown with a pink disc.}
	\label{fig:figure5}
\end{figure}
\end{landscape}

The first set of travelling wave solutions we report in  Figure \ref{fig:figure3} correspond to the simplest possible case where $D = \lambda = 1$ so that the skin and cancer cells are equally motile and equally proliferative.  Additional results are presented in Figures \ref{fig:figure4}-\ref{fig:figure5} for $D \ne 1$ and $\lambda = 1$, and for $D=1$ and $\lambda \ne 1$, respectively.  Results in Figure \ref{fig:figure4}-\ref{fig:figure5} are presented in the exact same format as in Figure \ref{fig:figure3} where we consider results for $c > 0$ and $c < 0$ separately in both cases.  In all cases we find that the travelling wave solutions from Equations (\ref{eq:PartialDiffUNonDim})-(\ref{eq:PartialDiffVNonDim}) compare very well with the physical portion of the phase plane trajectories and that the nonphysical portion of the phase plane trajectories do not play any role in the travelling wave solutions.

\subsection{Perturbation solution for $|c| \ll 1$} \label{sec:PerturbationSol}
All results in Figures \ref{fig:figure3}--\ref{fig:figure5} rely on  numerical solutions of Equations (\ref{eq:ODESysVXz})-(\ref{eq:XVsysdX}) to explore trajectories in the phase plane.   We now provide additional insight by constructing approximate perturbation solutions to complement these numerical explorations.  First we re-write Equations (\ref{eq:ODESysVXz})-(\ref{eq:XVsysdX}) as
\begin{equation}\label{eq:ODEXV}
\frac{\mathrm{d}X}{\mathrm{d}V} = \dfrac{-cX - \lambda V (1-V)}{D X},
\end{equation}
for which we seek a perturbation solution about $c=0$.  Substituting the expansion $X(V) =X_0 (V)+cX_1 (V)+ \mathcal{O}(c^2)$
leads to ordinary differential equations for $X_0 (V)$ and $X_1 (V)$ that can be solved exactly. With the initial condition $X(1)=0$, the two-term perturbation solution can be written as
\begin{equation}
\label{eq:VXPertubSol}
X(V) = \pm \sqrt{\frac{\lambda}{D}\left(- V^2 + \frac{2V^3}{3} + \frac{1}{3}\right)} - c \dfrac{(V-2)(1+2V)^{3/2}+\sqrt{27}}{5D(V-1)\sqrt{1+2V}} + \mathcal{O}(c^2).
\end{equation}
Retaining just the first term on the right of Equation (\ref{eq:VXPertubSol}) gives us an approximation that we refer to as an $\mathcal{O}(1)$ perturbation solution whereas retaining both terms on the right of Equation (\ref{eq:VXPertubSol}) gives us an approximation that we refer to as an $\mathcal{O}(c)$ perturbation solution.  We will now explore both these solutions.

\begin{figure}
		\centering
		\includegraphics[width=1\textwidth]{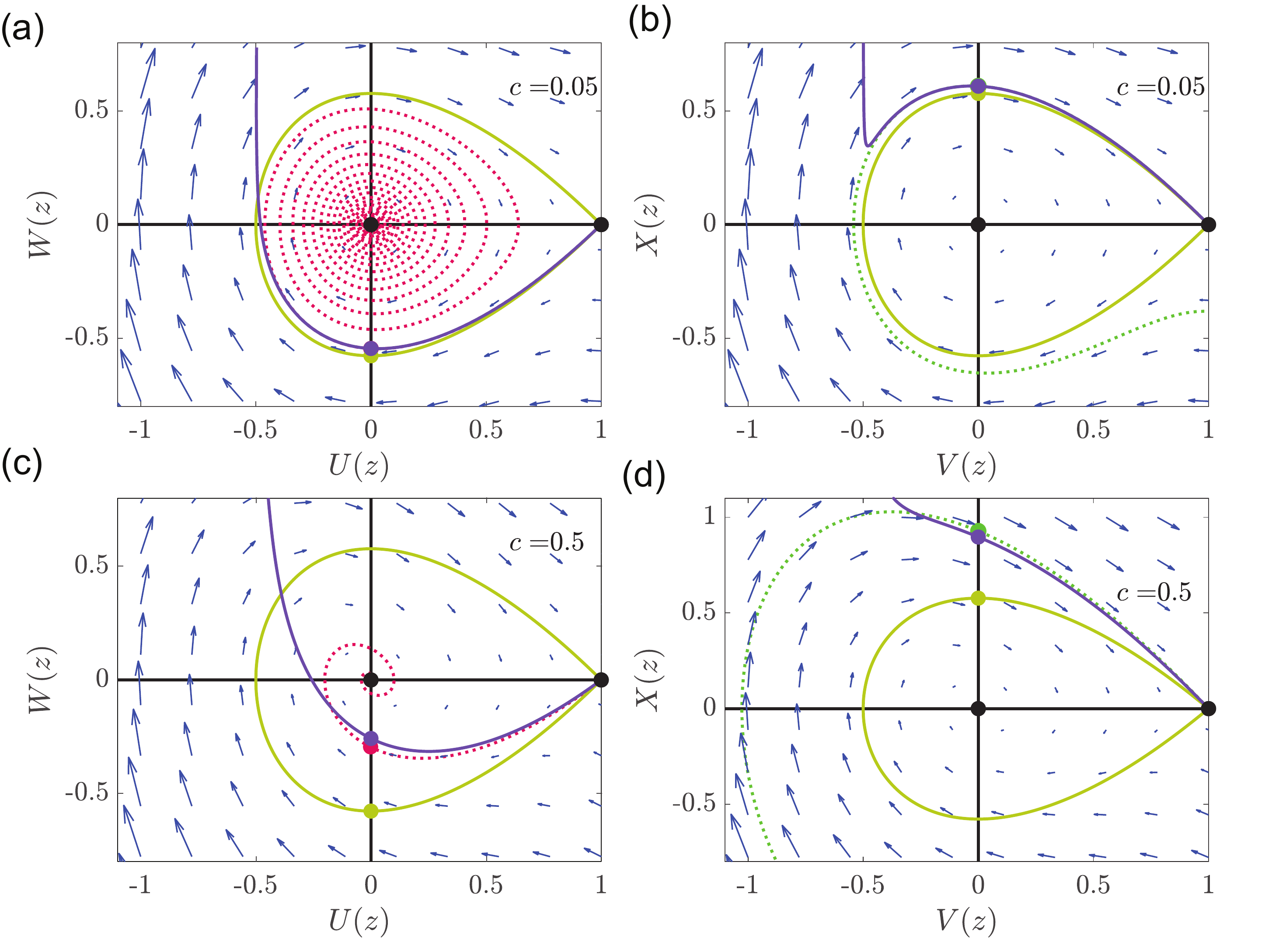}
		\caption{\textbf{Perturbation solution in the phase plane when $c > 0$ and $D = \lambda = 1$.}  Phase planes in (a)-(b), (c)-(d) compare numerical phase plane trajectories and perturbation solutions for $c=0.05$ and $c=0.5$, respectively.  Numerical estimates of the $U(W)$ and $V(X)$ trajectories are shown in dashed red and dashed green respectively.  The  $\mathcal{O}(1)$ and $\mathcal{O}(c)$ perturbation solutions are shown in solid yellow and solid purple, respectively.  Equilibrium points are shown with black discs.  The points at which the various solutions intersect the vertical axis are shown with various coloured discs corresponding to the colour of the particular trajectory. }
		\label{fig:figure6}
\end{figure}

Results in Figure \ref{fig:figure6}(a)-(b) show the $(U(z),W(z))$ and $(V(z),X(z))$ phase planes for $c=0.05$, respectively.  The numerically-generated trajectories are compared with both the $\mathcal{O}(1)$  and $\mathcal{O}(c)$ perturbation solutions.  Here we see that the   $\mathcal{O}(1)$ perturbation solution is a teardrop-shaped homoclinic trajectory to $(1,0)$.  In Figure \ref{fig:figure6}(a) we see that the $\mathcal{O}(1)$ perturbation solution is a reasonably accurate approximation of the numerical trajectory in the fourth quadrant for $U(W)$.  Similarly, in Figure \ref{fig:figure6}(b) we see that the $\mathcal{O}(1)$ perturbation solution is a reasonably accurate approximation of the numerical trajectory in the first quadrant for $V(X)$.   We also superimpose the $\mathcal{O}(c)$ perturbation solution in Figures \ref{fig:figure6}(a)-(b) but it is difficult to visually distinguish between the $\mathcal{O}(1)$ and $\mathcal{O}(c)$ solutions for $c=0.05$.

Results in Figure \ref{fig:figure6}(c)-(d) show the $(U(z),W(z))$ and $(V(z),X(z))$ phase planes for $c=0.5$, respectively.  In both cases we see that the $\mathcal{O}(1)$ perturbation solutions do not provide an accurate approximation of the numerical trajectories, whereas the  $\mathcal{O}(c)$ perturbation solutions compare very well with the physical part of the phase plane trajectories in both cases.   The comparison of the numerical phase plane trajectories and the perturbation solutions in Figure \ref{fig:figure6} is given for the most fundamental case where  $D = \lambda = 1$ and $c > 0$.  Additional comparisons for other choices of $D$, $\lambda$ and $c$ are provided in Appendix B.

\begin{figure}
	\centering
	\includegraphics[width=1\textwidth]{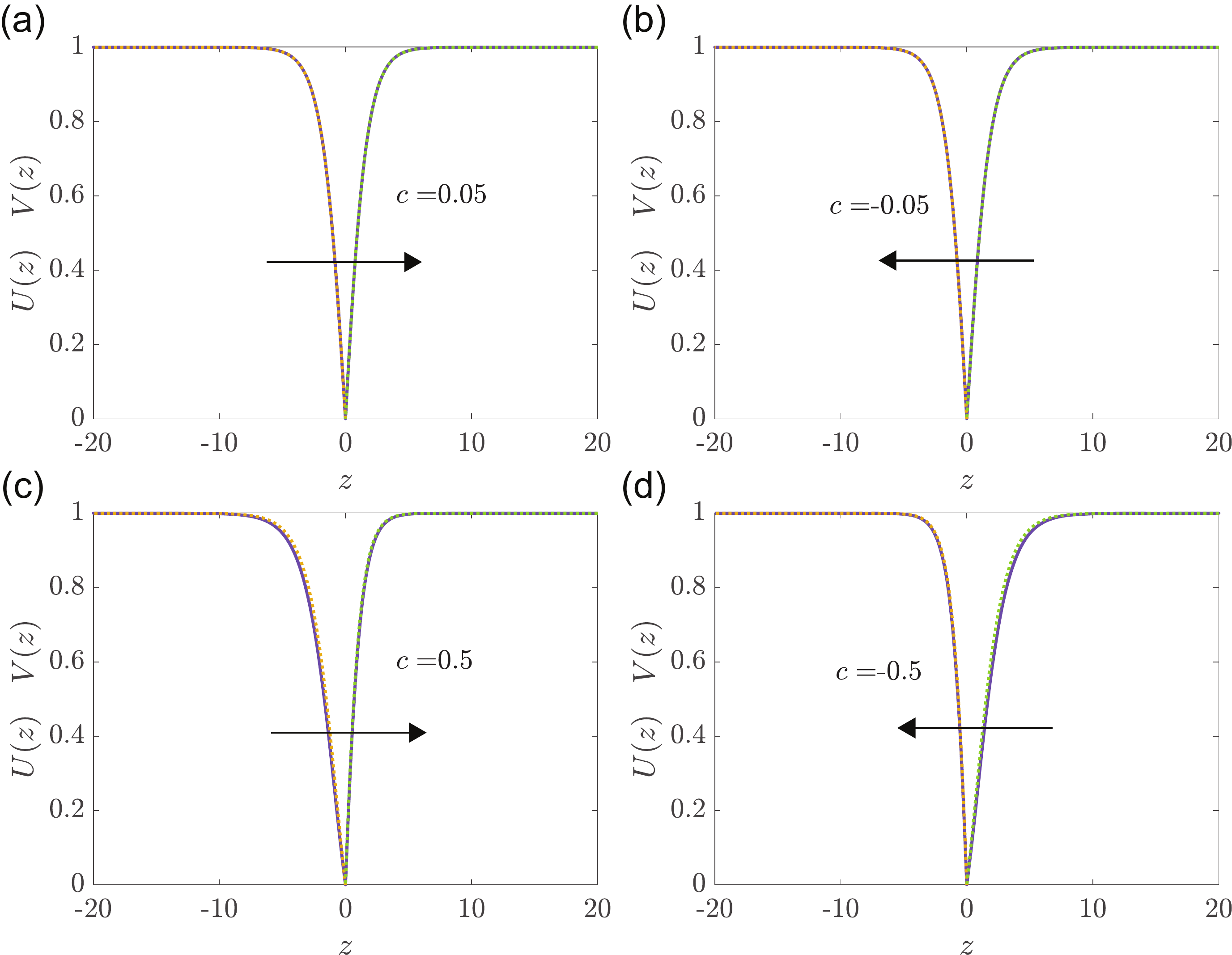}
	\caption{\textbf{Perturbation solution for the shape of the travelling waves for $D = \lambda = 1$.} Comparison of $U(z)$ and $V(z)$ from the $\mathcal{O}(c)$ perturbation solution (purple solid) with numerical estimates of the travelling wave obtained by solving Equations (\ref{eq:PartialDiffUNonDim})-(\ref{eq:PartialDiffVNonDim}) and shifting the profiles so that $U(0)=V(0)=0$.  Numerical estimates of $U(z)$ and $V(z)$ are shown in dashed yellow and dashed green lines, respectively.  Results are shown for: (a)--(b) $c= \pm 0.05$, and (c)--(d) $\pm c=-0.5$.}
	\label{fig:figure7}
\end{figure}

The comparison of the numerical phase plane trajectories with the perturbation solutions in Figure \ref{fig:figure6} shows the shape of $W(U)$ and $X(V)$ in the phase plane.  To explore how these solutions compare in the $z$ plane, we integrate both sides of Equation (\ref{eq:VXPertubSol}) with respect to $z$ numerically using a forward Euler approximation with constant step size, $\textrm{d}z=1 \times 10^{-4}$.  This numerical integration leads to estimates of the shape of the travelling waves that can be compared with the shapes of the travelling wave obtained from long-time numerical solutions of Equations (\ref{eq:PartialDiffUNonDim})-(\ref{eq:PartialDiffVNonDim}).   Figure \ref{fig:figure7} compares the shape of both $V(z)$ and $U(z)$ obtained from the $\mathcal{O}(c)$ perturbation solution with those obtained from  Equations (\ref{eq:PartialDiffUNonDim})-(\ref{eq:PartialDiffVNonDim}), where we see that the shape of both the $V(z)$ and $U(z)$ profiles compare extremely well for $c = \pm 0.05$, as expected.  It is also pleasing that the shape of the profiles compare quite well even for much larger values, $c = \pm 0.5$.  All results in Figure \ref{fig:figure7} correspond to the simplest case where $D = \lambda = 1$ and additional comparisons for other choices of $D$ and $\lambda$ and are provided in Appendix B.

\subsection{Qualitatively different long time behaviour} \label{sec:OtherSol}
All solutions in Figures \ref{fig:figure3}-\ref{fig:figure5} correspond particular choices of $u(x,0)$, $v(x,0)$, $\kappa_u$ and $\kappa_v$ that lead to long time travelling wave solutions.  However, we note that numerical simulations and more rigorous analysis of the simpler single-phase Fisher-Stefan moving boundary problems gives rise to a \textit{spreading-vanishing} dichotomy, whereby  certain initial conditions and parameter values lead to population extinction as $t \to \infty$~\cite{Elhachem2019,Simpson2020,Du2010,Du2011,Bunting2012,Du2012,Du2014a,Du2014b,Du2015}.  The main focus of our current work is to study travelling wave solutions since we are interested in situations where both populations are present, such as the images in Figure \ref{fig:figure1}(a)-(b).  To complement these solutions in Figure \ref{fig:figure3}--\ref{fig:figure7} we now briefly consider additional numerical solutions of Equations (\ref{eq:PartialDiffUNonDim})-(\ref{eq:PartialDiffVNonDim}) where similar extinction behaviour occurs in the two-phase problem.

\begin{figure}[H]
	\centering
	\includegraphics[width=1.0\textwidth]{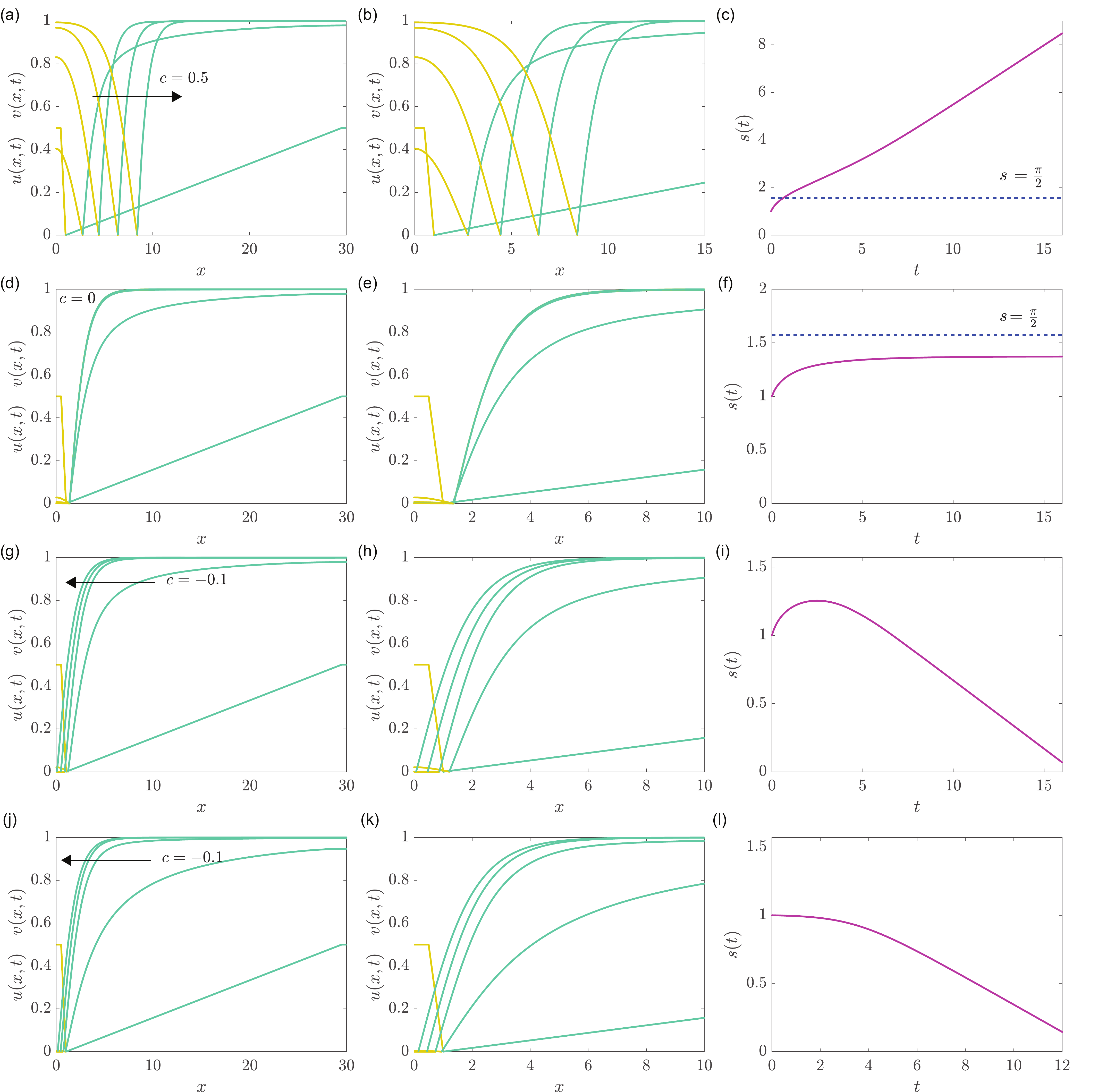}
	\caption{\textbf{Additional solutions with qualitatively different long time behaviour.} Four additional numerical solutions of Equations (\ref{eq:PartialDiffUNonDim})--(\ref{eq:PartialDiffVNonDim}).  Each solution corresponds to $D=1$, $\lambda=1$, with $L= 30$, $s(0) = 1$, $\beta = 0.5$, $\alpha = 0.5$.  Results in each row correspond to different values of $\kappa_u$ and $\kappa_v$: (a)--(c) corresponds to $\kappa_u = 2.2976$ and $\kappa_v =  0.1946$; (d)--(f) corresponds to $\kappa_u = 0.5$ and $\kappa_v = 0$; (g)--(i) corresponds to $\kappa_u = 0.5$ and $\kappa_v = 0.1946$; and (j)--(l) corresponds to $\kappa_u = 0.0001$ and $\kappa_v = 0.1946$.  The profiles in (a)-(b), (d)-(e) and (g)-(h) are shown at $t=0,4,8,12$ and $16$.  The density profiles in (i)-(j) are shown at $t=0,3,6,9$ and $12$. Profiles in the left column show the evolution of the solutions on $0 < x < 60$; profiles in the middle column show the details of these solutions on $0 < x < 10$, and profiles in the right column show the evolution of $s(t)$.}
	\label{fig:figure8}
\end{figure}

Figure \ref{fig:figure8} shows various results when we consider vary the initial condition and/or values of $\kappa_u$ and $\kappa_v$.  The first set of results in Figure \ref{fig:figure8}(a)--(c) shows a case in which $s(0)=1$. Here we see the solution evolving to travelling wave profile with positive speed of the type we have discussed in some detail. An important point to make here is that $s(0)< \pi/2$, which is a critical length in the corresponding one-phase problem~\cite{Elhachem2019,Simpson2020}. Based on the previously reported studies of the one-phase problem, our interpretation of the solution in Figure \ref{fig:figure8}(a)--(c) is that even though $s(0) < \pi/2$, travelling wave solutions are still possible provided the initial mass $\int _{0}^{s(0)} u(x,0) \, \textrm{d}x$ is sufficient to overcome mass lost at the moving boundary. On the other hand, in Figure \ref{fig:figure8}(d)--(f) the solution has the same parameter values as in Figure \ref{fig:figure8}(a)--(c), except that $\kappa_u$ and $\kappa_v$ are now reduced. In this case the moving boundary $x=s(t)$ moves to the right and approaches a steady state
value which is less than the critical length $\pi/2$, while the left population $u(x,t)$ goes extinct as $t \to \infty$. The extinction is caused by the fact that the rate at which mass associated with the $u(x,t)$ population is lost at $x=s(t)$ exceeds the rate at which the mass of $u(x,t)$ is gained by proliferation. These two examples are consistent with the spreading-vanishing dichotomy in the one-phase problem~\cite{Elhachem2019,Simpson2020}.

\newpage
Additional results in Figures \ref{fig:figure8}(g)--(i) and (j)--(l) show two further solutions with different choices of $\kappa_u$ and $\kappa_v$.  In both these cases we see that the $u(x,t)$ profile eventually becomes extinct whereas the $v(x,t)$ profile eventually forms a travelling wave solution with $c=-0.1$.  Subtle differences, highlighted in Figure \ref{fig:figure8}(i) and (l), show the temporal behaviour in terms of the movement of the interface, $s(t)$.  The case in Figure \ref{fig:figure8}(l) leads to a monotonically decreasing $s(t)$, whereas the case in Figure \ref{fig:figure8}(i) leads to $s(t)$ that is initially increasing before eventually decreasing at later time.  This kind of nonmonotone behaviour of $s(t)$ is very interesting because the standard single phase Fisher-Stefan model appears to only lead to monotone $s(t)$, whereas our two-phase analogue leads to more interesting and nuanced behaviours.

\section{Conclusion} \label{sec:Conclusion}

In this work we consider a novel mathematical model of cell invasion which takes the form of a two-phase moving boundary problem.  This modelling strategy is both biologically relevant and mathematically novel.  The moving boundary model leads to travelling wave solutions with a clearly defined moving front.  This is advantageous over the classical Fisher--Kolmogorov model and extensions because travelling wave solutions of those models do not have this property.  From a biological point of view, our model describes the migration and proliferation of two populations of cells, $u(x,t)$ and $v(x,t)$, and this allows us to model a population of cancer cells, $u(x,t)$, invading into a population of surrounding cells, $v(x,t)$.  This scenario is relevant to melanoma cells invading into surrounding skin cells, as shown in Figure \ref{fig:figure1}(a)-(b).  Interestingly, the moving boundary model leads to travelling wave solutions that move in either the positive or negative direction, meaning that we can simulate malignant invasion as well as malignant retreat.  This is very different to travelling wave solutions of the Fisher-Kolmogorov and Porous-Fisher models because those models only ever predict malignant advance and never predict malignant retreat.

The two-phase moving boundary model is also very interesting mathematically.  In this work we analyse travelling wave solutions where we show that the $U(z) = u(x-ct)$ and $V(z)=v(x-ct)$ travelling waves can be analysed in two separate phase planes.  These two phase planes are identical to the phase plane that arises in the classical analysis of travelling wave solutions of the Fisher-Kolmogorov model.  This phase plane contains two equilibria: (i) $(1,0)$ is a saddle for all $c$; and (ii) $(0,0)$ is a stable node if $c \ge 2\sqrt{\lambda D}$ or a stable spiral for $c < 2\sqrt{\lambda D}$.  Normally, in the case of travelling waves solutions of the Fisher--Kolmogorov model we are interested in a heteroclinic trajectory between these two equilibria and so we require $c \ge 2 \sqrt{\lambda D}$ to avoid the nonphysical negative densities that arise from spirals in the phase plane.  In contrast, travelling wave solutions of our moving boundary model have $c < 2 \sqrt{\lambda D}$ and so these normally-discarded trajectories turn out to be very useful.

For our two-phase moving boundary model we use numerical simulations and perturbation methods to confirm that the travelling wave solutions for  $U(z)$ and $V(z)$ are associated with trajectories in the classical Fisher--Kolmogorov phase plane that are normally disregarded as being nonphysical.  In the cases we consider with $c > 0$, the $U(z)$ travelling wave is associated with the heteroclinic trajectory that leaves $(1,0)$ along the unstable manifold and eventually spiralling into $(0,0)$. Here we have the restriction that the travelling wave solution is only associated with the first part of that trajectory where $U(z) > 0$. Similarly, the $V(z)$ travelling wave is associated with the trajectory that approaches $(1,0)$ along the stable manifold.  Here we have the restriction that the travelling wave is associated with that part of the trajectory near $(1,0)$ where $V(z) > 0$.   For travelling wave solutions with $c < 0$ it is the other way around:  the $U(z)$ travelling wave is associated with the trajectory that eventually moves into $(1,0)$ from infinity, whereas the $V(z)$ travelling wave is associated with the trajectory that eventually spirals into $(0,0)$.  It is very interesting that both these trajectories come from the phase plane for the well--studied Fisher--Kolmogorov equation, except that these trajectories are not normally considered in any detail.

There are many ways that our model could be extended to incorporate additional features.  For example, from a practical point of view, all work presented here involves applying these models in a standard one-dimensional Cartesian geometry and it would also be interesting to apply these models in a radial geometry to study the outward invasion of a spherical tumour or the closure of a disc-shaped wound~\cite{Treloar2014}.  Further considerations could be to explicitly model how malignant cells produce proteases and other chemical signals and to explore how such signals can be incorporated into the evolution equation for the moving boundary~\cite{Smallbone2005,Holder2014}.   From a more mathematical point of view, additional questions of interest are to precisely study under which conditions solutions go to travelling waves or become extinct, and to study the limit $t \to \infty$ with care to determine how quickly travelling wave solutions develop.

\vspace{1.0cm}
\noindent
\textit{Acknowledgements}. This work is supported by the Australian Research Council (DP200100177, DP190103757). We thank Dr Wang Jin for advice on the numerical algorithms, and we thank Emeritus Professor Sean McElwain for his feedback.\\

\noindent
\textit{Contributions}.  All authors conceived and designed the study; M.El-H. performed numerical calculations.  All authors
drafted the article and gave final approval for publication. \\

\noindent
\textit{Competing Interests}. We have no competing interests.\\

\newpage
\section{Appendix A: Numerical Methods}
\label{AppA}
Liberally commented MATLAB implementations of all numerical algorithms used to generate the solutions of the differential equations in this work are available on \href{https://github.com/ProfMJSimpson/El-Hachem2020}{GitHub}.

\subsection{Partial differential equation} \label{AppAPDE}

As we explained in the main document, the partial differential equation models are transformed to a fixed domain, Equations (16)-(17) on $0 \le \xi \le 1$ and $1 \le \eta \le 2$, respectively.  To solve these transformed partial differential equations we discretize the $\xi$ and $\eta$ domains uniformly.  In principle we use $m$ equally-spaced mesh points for $\xi$, $m = 1/\Delta \xi +1$, and $n$ equally-spaced mesh points for $\eta$, $n = 1/\Delta \eta+1$.  In practice we usually implement the numerical solution with $m=n$ by setting $\Delta \xi = \Delta \eta$.  This is convenient, but not necessary.

Using a central difference approximation for the transformed spatial variable and an implicit Euler approximation for the temporal derivatives~\cite{Simpson2005,Simpson2007b}, at the central nodes on both meshes we have
\begin{align}
\label{eq:PartialUDiscrete}
\begin{split}
\frac{u_{i}^{j+1}-u_{i}^{j}}{\Delta t} &= \frac{\left(u_{i+1}^{j+1}-2 u_{i}^{j+1}+u_{i-1}^{j+1}\right)}{ (s^{j+1} \Delta \xi)^{2}}+
\frac{\xi_i \left(s^{j+1}-s^{j}\right)\left(u_{i+1}^{j+1}-u_{i-1}^{j+1}\right)}{2 s^{j+1} \Delta t \Delta \xi}\\
&+ u_{i}^{j+1}\left(1-u_{i}^{j+1}\right), \quad i = 2, \ldots, m-1,
\end{split}
\\
\begin{split}
\frac{v_{i}^{j+1}-v_{i}^{j}}{\Delta t}  &= \frac{D \left(v_{i+1}^{j+1}-2 v_{i}^{j+1}+v_{i-1}^{j+1}\right)}{ \left( \left(L-s^{j+1}\right)\Delta \eta\right)^{2}} +    \frac{\left(2 - \eta_i\right)\left(v_{i+1}^{j+1}-v_{i-1}^{j+1}\right)\left(s^{j+1}-s^{j}\right)}{2 \Delta t \Delta \eta \left(L-s^{j+1}\right)}\\
&+ \lambda
v_{i}^{j+1}\left(1-v_{i}^{j+1}\right), \quad i = 2, \ldots, n-1,
\end{split}
\label{eq:PartialVDiscrete}
\end{align}
where the subscript $i$ denotes the mesh point and the superscript $j$ denotes the time, where $t = j \Delta t$.

To enforce the boundary conditions we set $\partial u / \partial \xi=0$ at $\xi=0$ and $\partial v /\partial \eta = 0$ at $\eta=2$, further we set $u = v = 0$ at the moving boundary where $\xi=\eta=1$, leading to
\begin{equation}
\label{eq:NeumannFD}
u_{2}^{j+1}= u_{1}^{j+1}, \quad v_{n}^{j+1} = v_{n-1}^{j+1}, \quad u_{m}^{j+1} = v_{1}^{j+1} = 0.
\end{equation}

To advance the discrete system from time $t$ to $t + \Delta t$ we solve the system of nonlinear algebraic equations, Equations (\ref{eq:PartialUDiscrete})-(\ref{eq:NeumannFD}), using Newton-Raphson iteration~\cite{Burden2011}. During each iteration of the Newton-Raphson algorithm we estimate the position of the moving boundary using the discretised Stefan condition,
\begin{equation}
\frac{s^{j+1} - s^{j}}{\Delta t } =  -\kappa_{u}\frac{u_{m}^{j+1}-u_{m-1}^{j+1}}{ s^{j+1} \Delta \xi} - \kappa_{v}\frac{v_{2}^{j+1}-v_{1}^{j+1}}{\left(L-s^{j+1}\right) \Delta \eta}.
\label{eq:Lupdatediscretise}
\end{equation}
Within each time step the Newton-Raphson iterations continue until the maximum chance in the dependent variables is less than the tolerance  $\epsilon$.  All results in this work are obtained by setting $\epsilon = 1 \times 10^{-8}$, $\Delta \xi = \Delta \eta = 2.5 \times 10^{-4}$ and $\Delta t = 1 \times 10^{-3}$, and we find that these values are sufficient to produce grid-independent results.  However, we recommend that care be taken when using the algorithms on \href{https://github.com/ProfMJSimpson/El-Hachem2020}{GitHub} for different choices of parameters, especially when considering larger values of $\kappa_u$ and $\kappa_v$, which can require a much denser mesh to give grid-independent results.

\subsection{Phase plane}\label{AppAPhasePlane}
To construct the phase planes we solve Equations (26)--(27) numerically using Heun's method with a constant step size $\textrm{d}z$. In most cases we are interested in examining trajectories that either enter or leave the saddle $(1,0)$ along the stable or unstable manifold, respectively  Therefore, it is important that the initial condition we chose when solving Equations (26)--(27) are on the appropriate stable or unstable manifold and sufficiently close to $(1,0)$.  To choose this point we use the MATLAB \textit{eig} function~\cite{eig} to calculate the eigenvalues and eigenvectors for the particular choice of $c$, $D$ and $\lambda$ of interest. The flow of the dynamical system are plotted on the phase planes using the MATLAB \textit{quiver} function~\cite{quiver}.

\newpage

\section{Appendix B: Additional results}\label{AppB}
Results in Figure 6 are presented for $D = \lambda = 1$ and $c > 0$ only.  Similarly, results in Figure 7 are presented for $D = \lambda = 1$ only.  Here, in Section \ref{AppBPhasePlane} we present additional phase plane results where $D \ne 1$, $\lambda \ne 1$ and $c < 0$.  Similarly, here in Section \ref{AppBPhysicalPlane} we present additional results where we plot $U(z)$ and $V(z)$ where $D \ne 1$ and $\lambda \ne 1$.  In all cases we have a good match between the perturbation solutions and numerical solutions provided that the wavespeed is sufficiently close to zero, as expected.

\subsection{Additional perturbation results in the phase plane}\label{AppBPhasePlane}

\begin{figure}[H]
	\centering
	\includegraphics[width=1\textwidth]{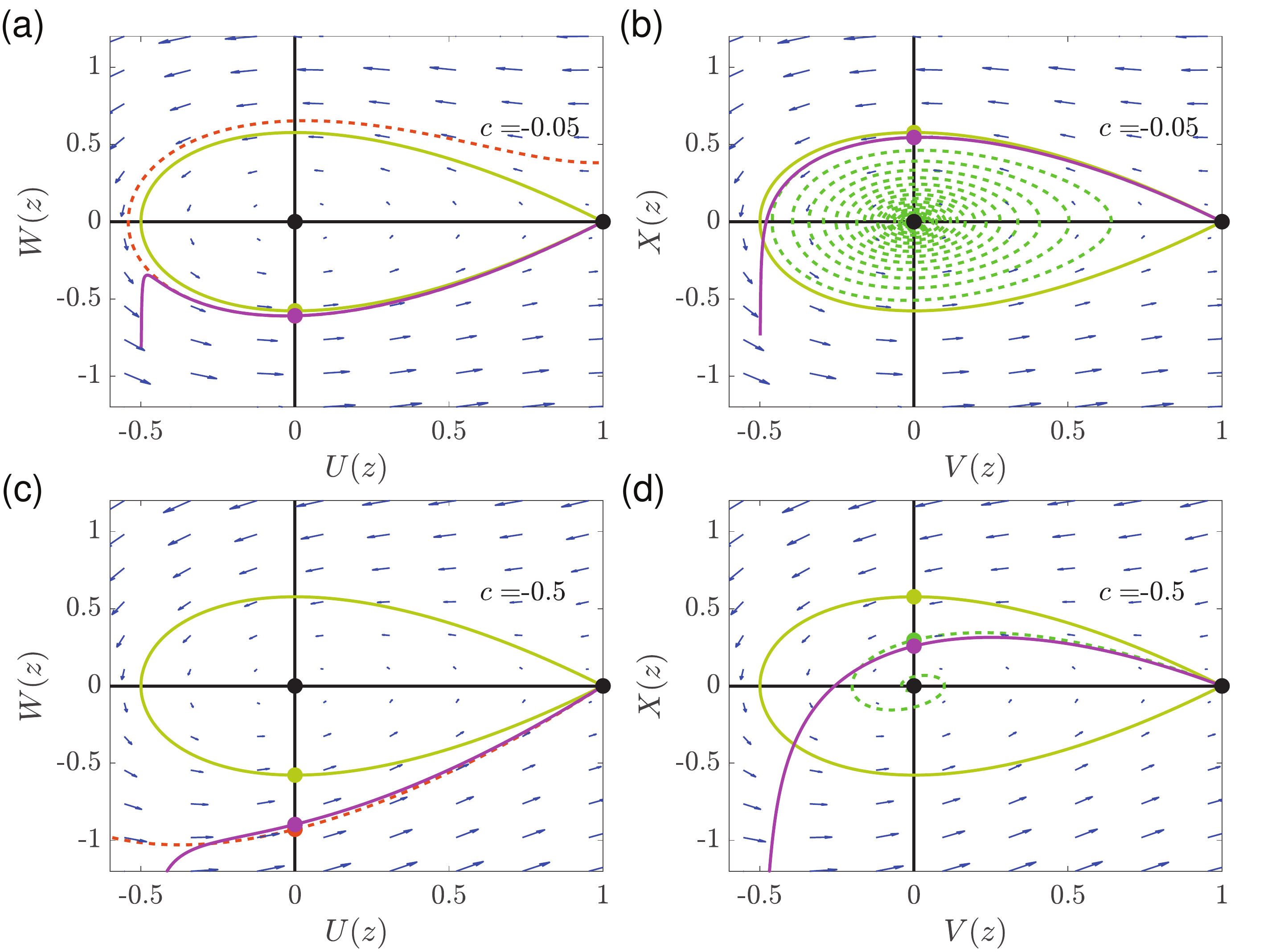}
	\caption{\textbf{Perturbation solution for the phase plane trajectories when $c < 0$ and $D = \lambda = 1$.}  Phase planes in (a)-(b), (c)-(d) compare numerical phase plane trajectories and perturbation solutions for $c=-0.05$ and $c=-0.5$, respectively.  Numerical estimates of the $U(W)$ and $V(X)$ trajectories are shown in dashed red and dashed green respectively.  The  $\mathcal{O}(1)$ and $\mathcal{O}(c)$ perturbation solutions are shown in solid yellow and solid purple, respectively.  Equilibrium points are shown with black discs.  The points at which the various solutions intersect the vertical axis are shown with various coloured discs corresponding to the colour of the particular trajectory.}
	\label{fig:figS1}
\end{figure}

\begin{figure}[H]
	\centering
	\includegraphics[width=1\textwidth]{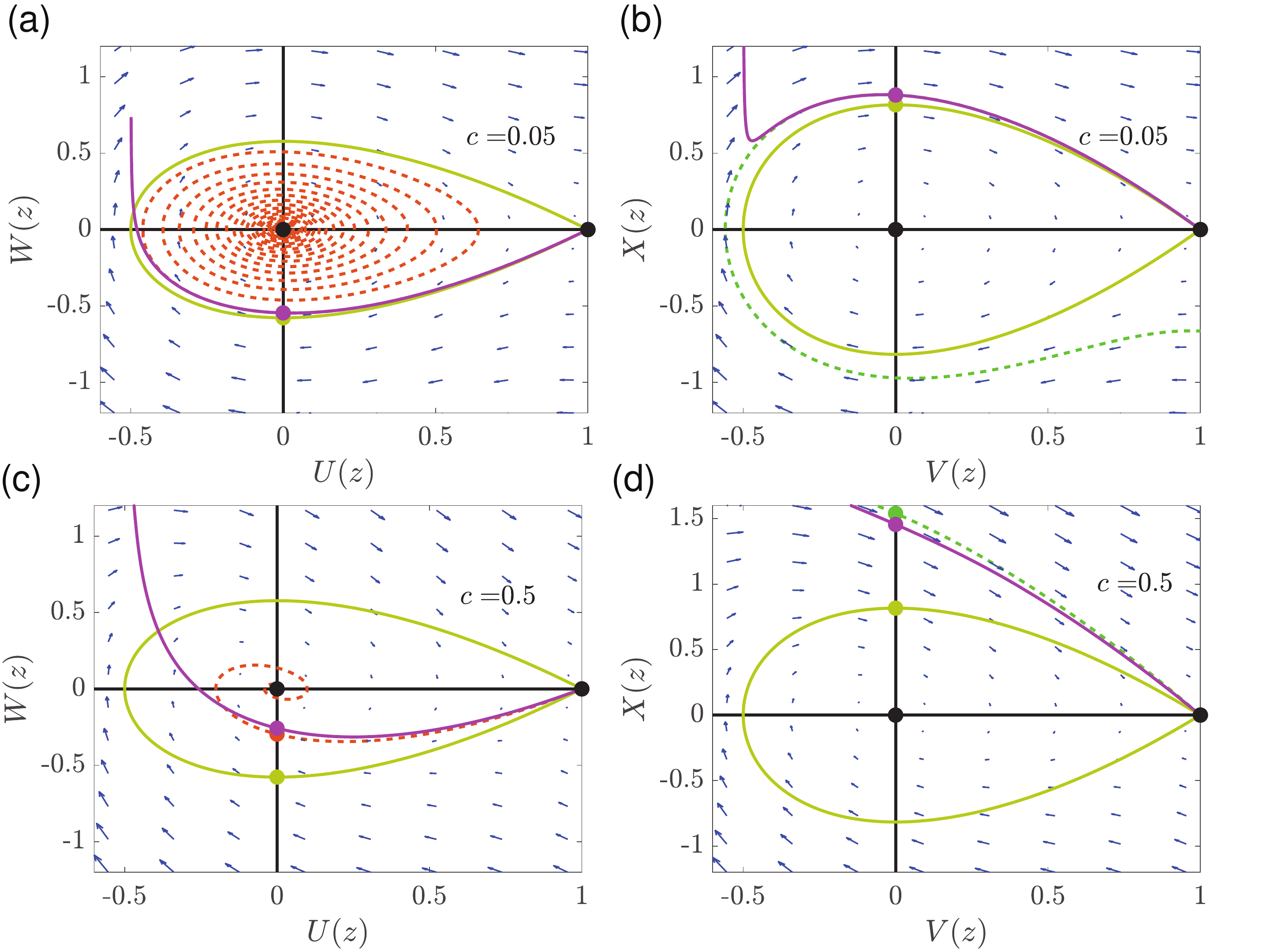}
	\caption{\textbf{Perturbation solution for the phase plane trajectories when $c > 0$, $D = 0.5$ and $\lambda = 1$.}  Phase planes in (a)-(b), (c)-(d) compare numerical phase plane trajectories and perturbation solutions for $c=0.05$ and $c=0.5$, respectively.  Numerical estimates of the $U(W)$ and $V(X)$ trajectories are shown in dashed red and dashed green respectively.  The  $\mathcal{O}(1)$ and $\mathcal{O}(c)$ perturbation solutions are shown in solid yellow and solid purple, respectively.  Equilibrium points are shown with black discs.  The points at which the various solutions intersect the vertical axis are shown with various coloured discs corresponding to the colour of the particular trajectory.}
	\label{fig:figS2}
\end{figure}

\begin{figure}[H]
	\centering
	\includegraphics[width=1\textwidth]{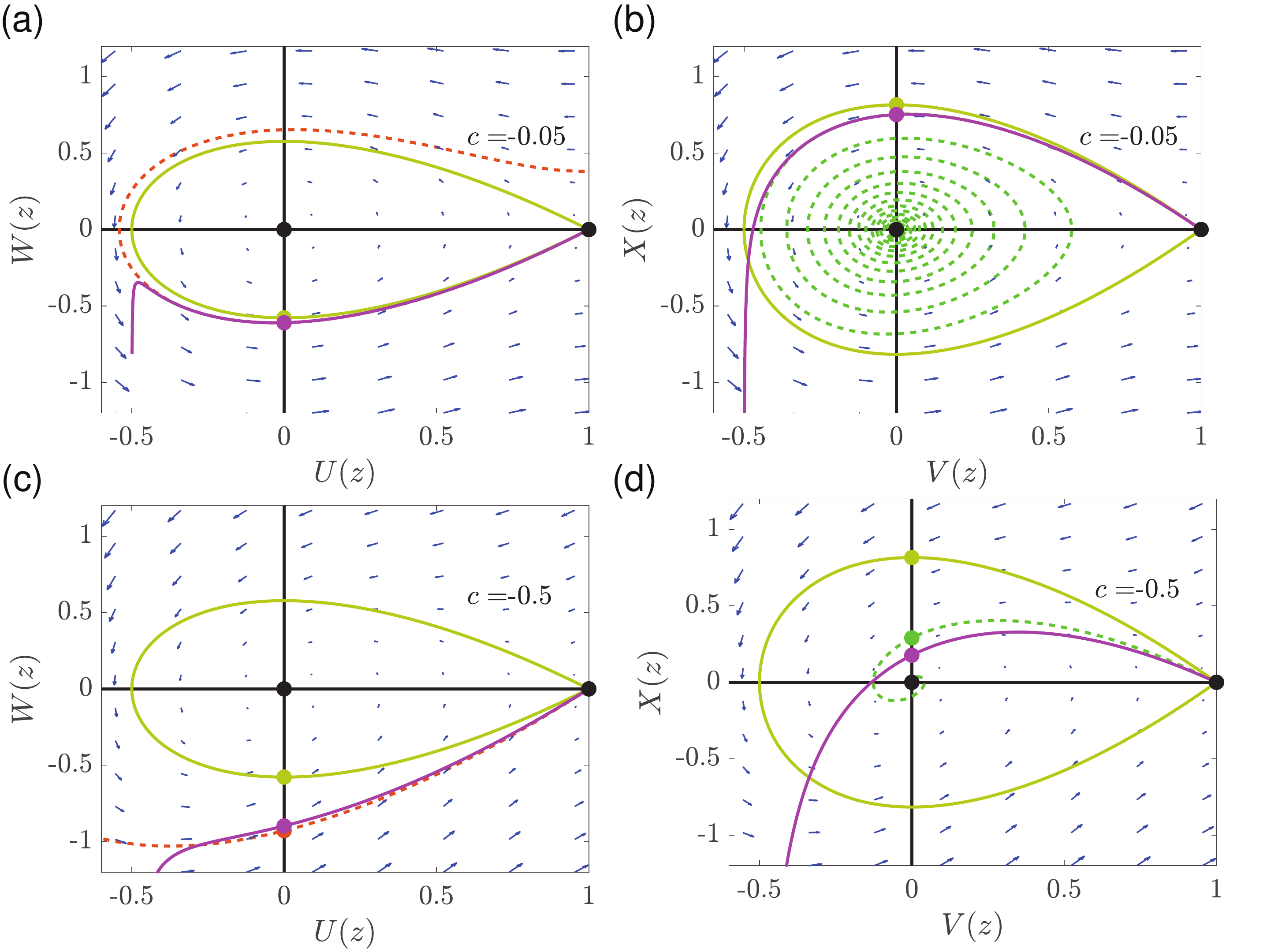}
	\caption{\textbf{Perturbation solution for the phase plane trajectories when $c < 0$, $D = 0.5$ and $\lambda = 1$.}  Phase planes in (a)-(b), (c)-(d) compare numerical phase plane trajectories and perturbation solutions for $c=-0.05$ and $c=-0.5$, respectively.  Numerical estimates of the $U(W)$ and $V(X)$ trajectories are shown in dashed red and dashed green respectively.  The  $\mathcal{O}(1)$ and $\mathcal{O}(c)$ perturbation solutions are shown in solid yellow and solid purple, respectively.  Equilibrium points are shown with black discs.  The points at which the various solutions intersect the vertical axis are shown with various coloured discs corresponding to the colour of the particular trajectory.}
	\label{fig:figS3}
\end{figure}

\begin{figure}[H]
	\centering
	\includegraphics[width=1\textwidth]{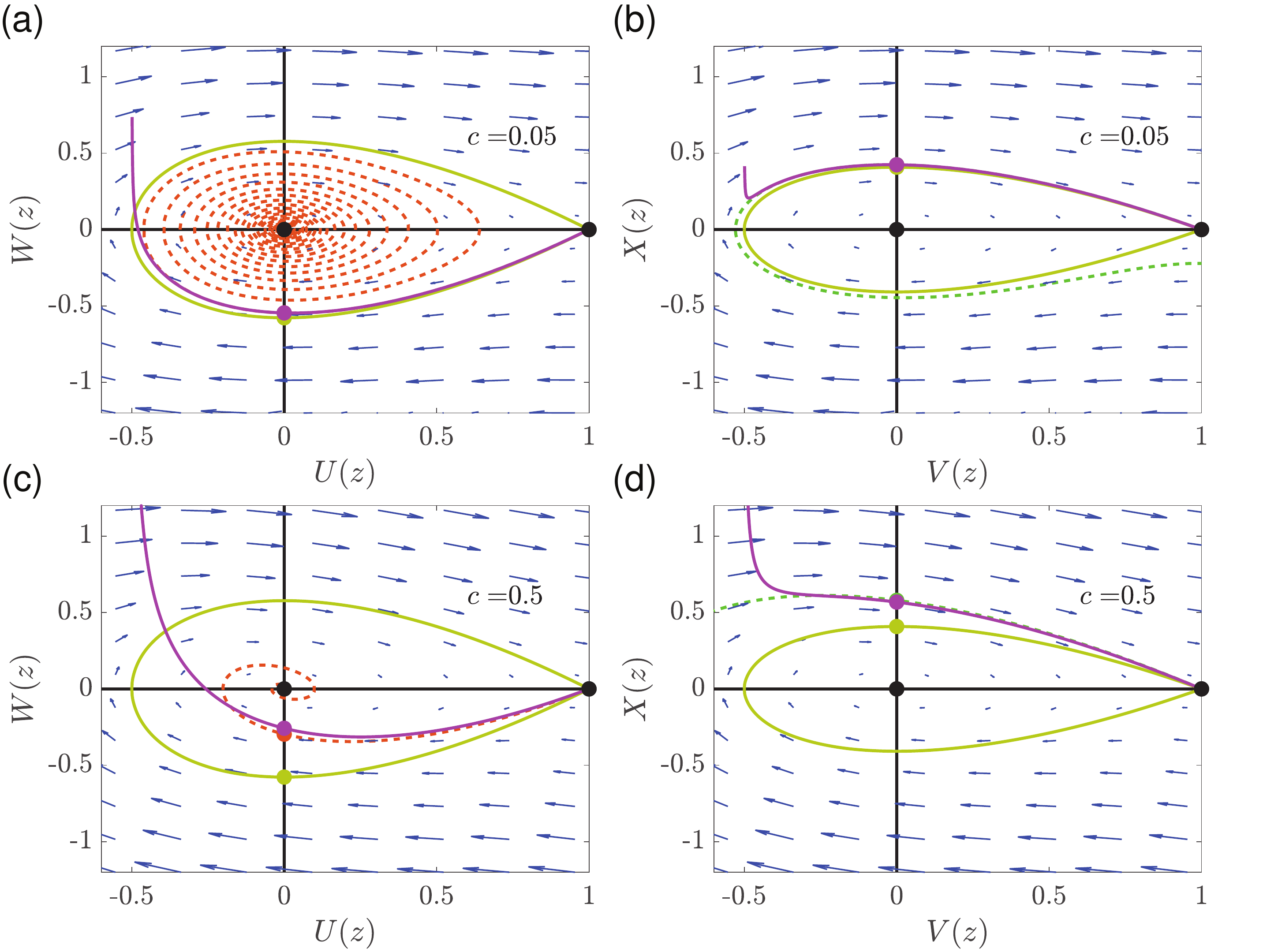}
	\caption{\textbf{Perturbation solution for the phase plane trajectories when $c > 0$, $D = 2$ and $\lambda = 1$.}  Phase planes in (a)-(b), (c)-(d) compare numerical phase plane trajectories and perturbation solutions for $c=0.05$ and $c=0.5$, respectively.  Numerical estimates of the $U(W)$ and $V(X)$ trajectories are shown in dashed red and dashed green respectively.  The  $\mathcal{O}(1)$ and $\mathcal{O}(c)$ perturbation solutions are shown in solid yellow and solid purple, respectively.  Equilibrium points are shown with black discs.  The points at which the various solutions intersect the vertical axis are shown with various coloured discs corresponding to the colour of the particular trajectory.}
	\label{fig:figS4}
\end{figure}

\begin{figure}
	\centering
	\includegraphics[width=1\textwidth]{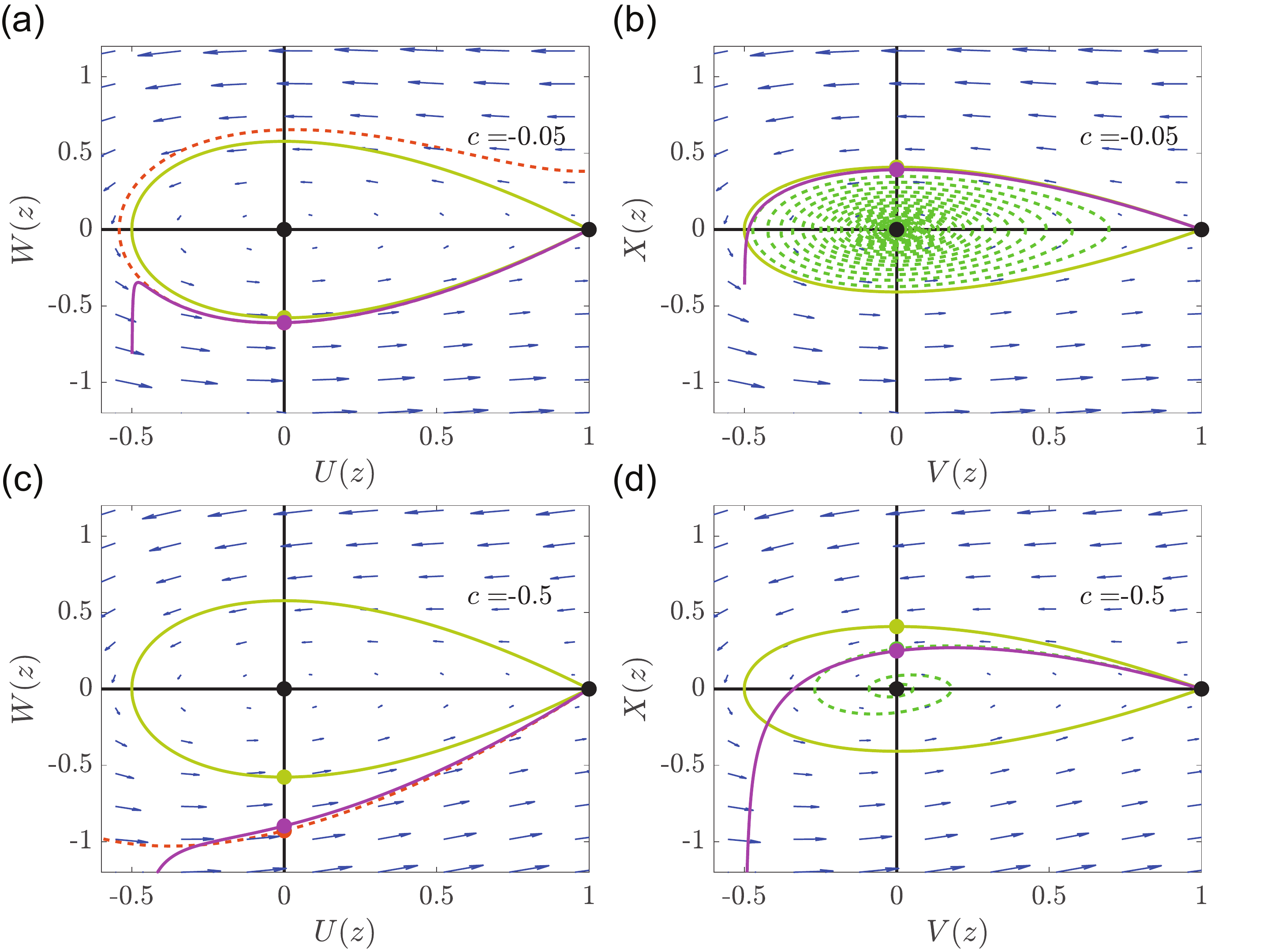}
	\caption{\textbf{Perturbation solution for the phase plane trajectories when $c < 0$, $D = 2$ and $\lambda = 1$.}  Phase planes in (a)-(b), (c)-(d) compare numerical phase plane trajectories and perturbation solutions for $c=-0.05$ and $c=-0.5$, respectively.  Numerical estimates of the $U(W)$ and $V(X)$ trajectories are shown in dashed red and dashed green respectively.  The  $\mathcal{O}(1)$ and $\mathcal{O}(c)$ perturbation solutions are shown in solid yellow and solid purple, respectively.  Equilibrium points are shown with black discs.  The points at which the various solutions intersect the vertical axis are shown with various coloured discs corresponding to the colour of the particular trajectory.}
	\label{fig:figS5}
\end{figure}

\begin{figure}[H]
	\centering
	\includegraphics[width=1\textwidth]{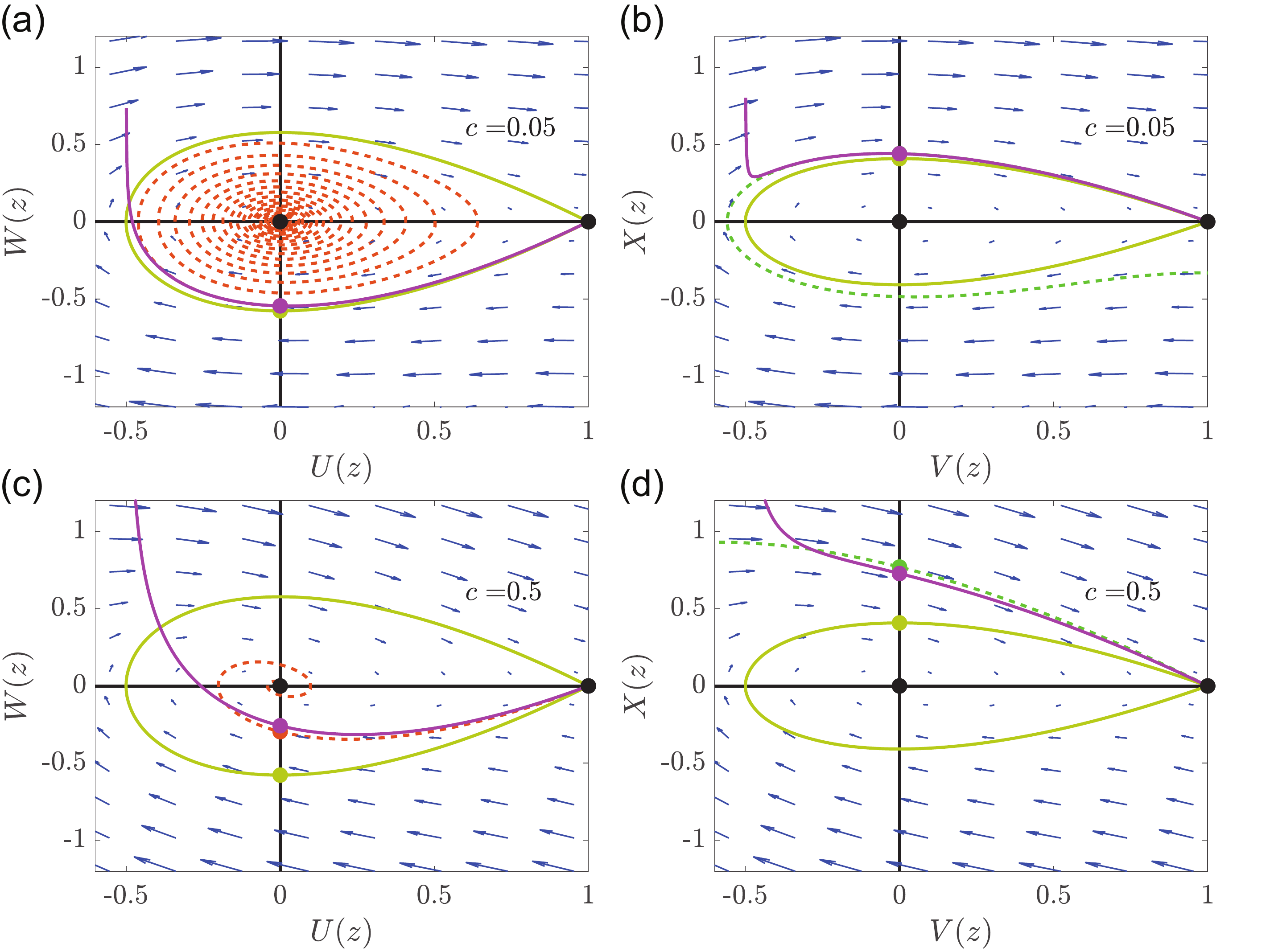}
	\caption{\textbf{Perturbation solution for the phase plane trajectories when $c > 0$, $D = 1$ and $\lambda = 0.5$.}  Phase planes in (a)-(b), (c)-(d) compare numerical phase plane trajectories and perturbation solutions for $c=0.05$ and $c=0.5$, respectively.  Numerical estimates of the $U(W)$ and $V(X)$ trajectories are shown in dashed red and dashed green respectively.  The  $\mathcal{O}(1)$ and $\mathcal{O}(c)$ perturbation solutions are shown in solid yellow and solid purple, respectively.  Equilibrium points are shown with black discs.  The points at which the various solutions intersect the vertical axis are shown with various coloured discs corresponding to the colour of the particular trajectory.}
	\label{fig:figS6}
\end{figure}

\begin{figure}[H]
	\centering
	\includegraphics[width=1\textwidth]{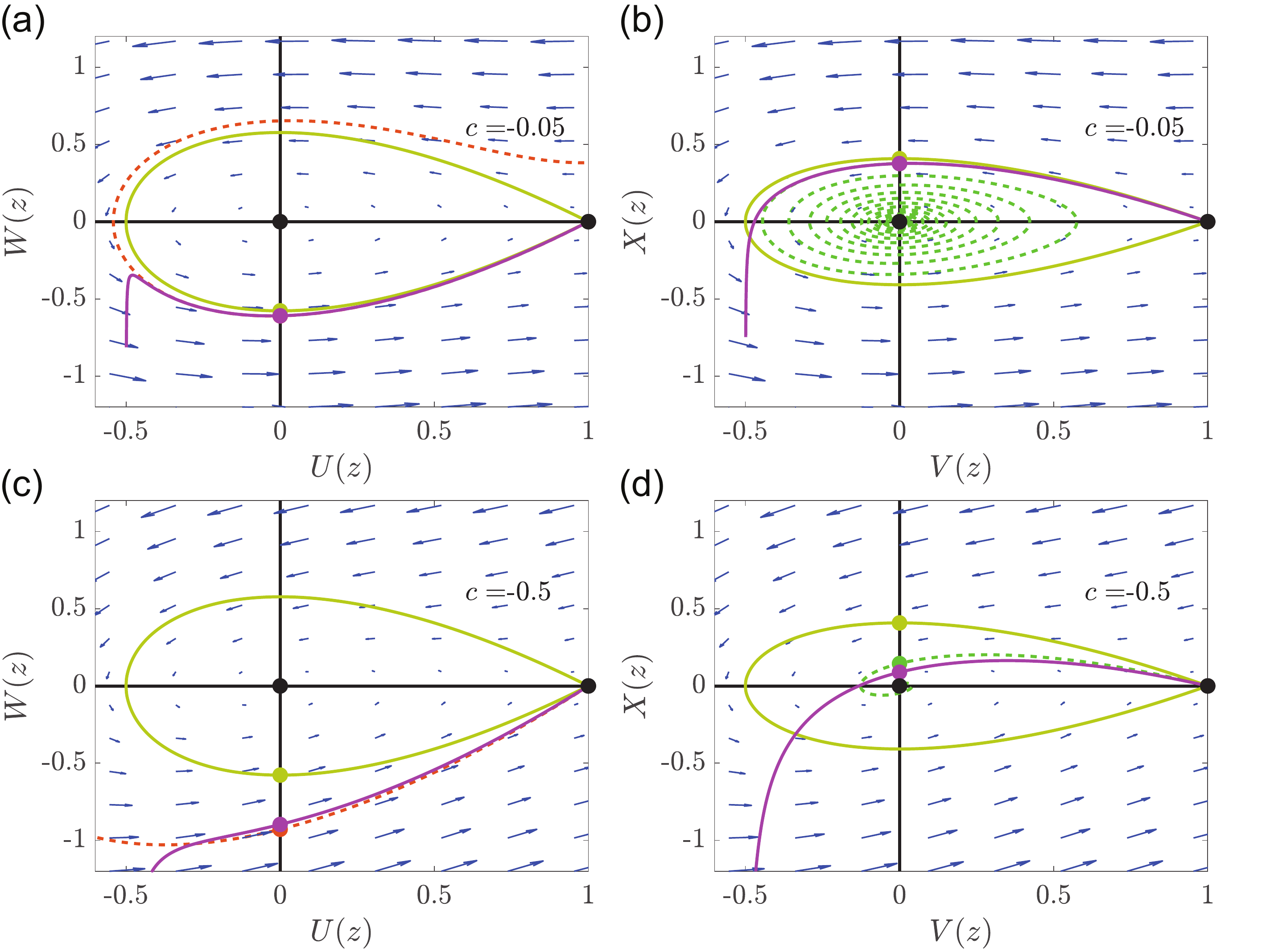}
	\caption{\textbf{Perturbation solution for the phase plane trajectories when $c < 0$, $D = 1$ and $\lambda = 0.5$.}  Phase planes in (a)-(b), (c)-(d) compare numerical phase plane trajectories and perturbation solutions for $c=-0.05$ and $c=-0.5$, respectively.  Numerical estimates of the $U(W)$ and $V(X)$ trajectories are shown in dashed red and dashed green respectively.  The  $\mathcal{O}(1)$ and $\mathcal{O}(c)$ perturbation solutions are shown in solid yellow and solid purple, respectively.  Equilibrium points are shown with black discs.  The points at which the various solutions intersect the vertical axis are shown with various coloured discs corresponding to the colour of the particular trajectory.}
	\label{fig:figS7}
\end{figure}

\begin{figure}[H]
	\centering
	\includegraphics[width=1\textwidth]{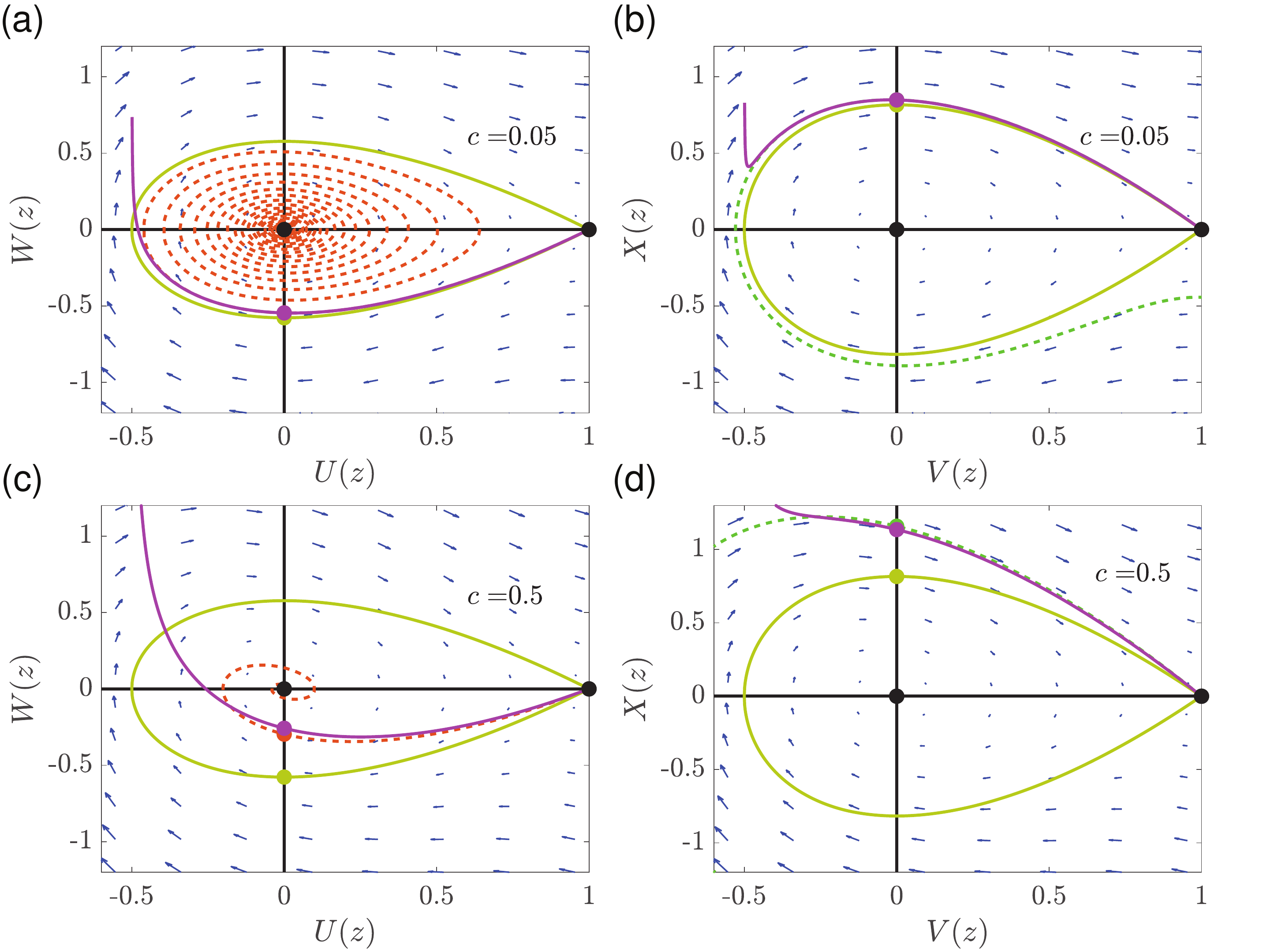}
	\caption{\textbf{Perturbation solution for the phase plane trajectories when $c > 0$, $D = 1$ and $\lambda = 2$.}  Phase planes in (a)-(b), (c)-(d) compare numerical phase plane trajectories and perturbation solutions for $c=0.05$ and $c=0.5$, respectively.  Numerical estimates of the $U(W)$ and $V(X)$ trajectories are shown in dashed red and dashed green respectively.  The  $\mathcal{O}(1)$ and $\mathcal{O}(c)$ perturbation solutions are shown in solid yellow and solid purple, respectively.  Equilibrium points are shown with black discs.  The points at which the various solutions intersect the vertical axis are shown with various coloured discs corresponding to the colour of the particular trajectory.}
	\label{fig:figS8}
\end{figure}

\begin{figure}[H]
	\centering
	\includegraphics[width=1\textwidth]{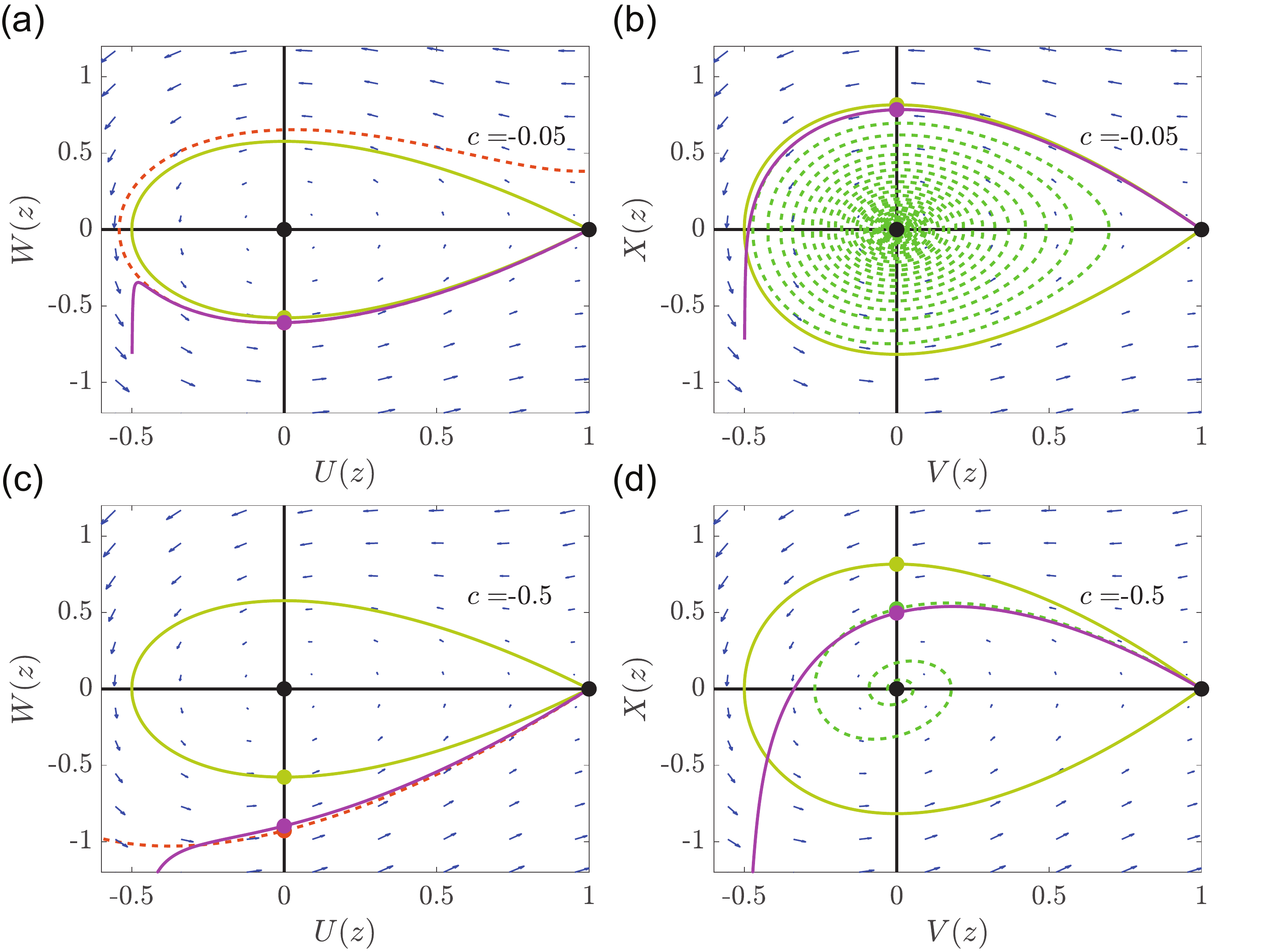}
	\caption{\textbf{Perturbation solution for the phase plane trajectories when $c < 0$, $D = 1$ and $\lambda = 2$.}  Phase planes in (a)-(b), (c)-(d) compare numerical phase plane trajectories and perturbation solutions for $c=-0.05$ and $c=-0.5$, respectively.  Numerical estimates of the $U(W)$ and $V(X)$ trajectories are shown in dashed red and dashed green respectively.  The  $\mathcal{O}(1)$ and $\mathcal{O}(c)$ perturbation solutions are shown in solid yellow and solid purple, respectively.  Equilibrium points are shown with black discs.  The points at which the various solutions intersect the vertical axis are shown with various coloured discs corresponding to the colour of the particular trajectory.}
	\label{fig:figS9}
\end{figure}

\subsection{Additional perturbation results presented in the $z$ coordinate}\label{AppBPhysicalPlane}

\begin{figure}[H]
	\centering
	\includegraphics[width=1\textwidth]{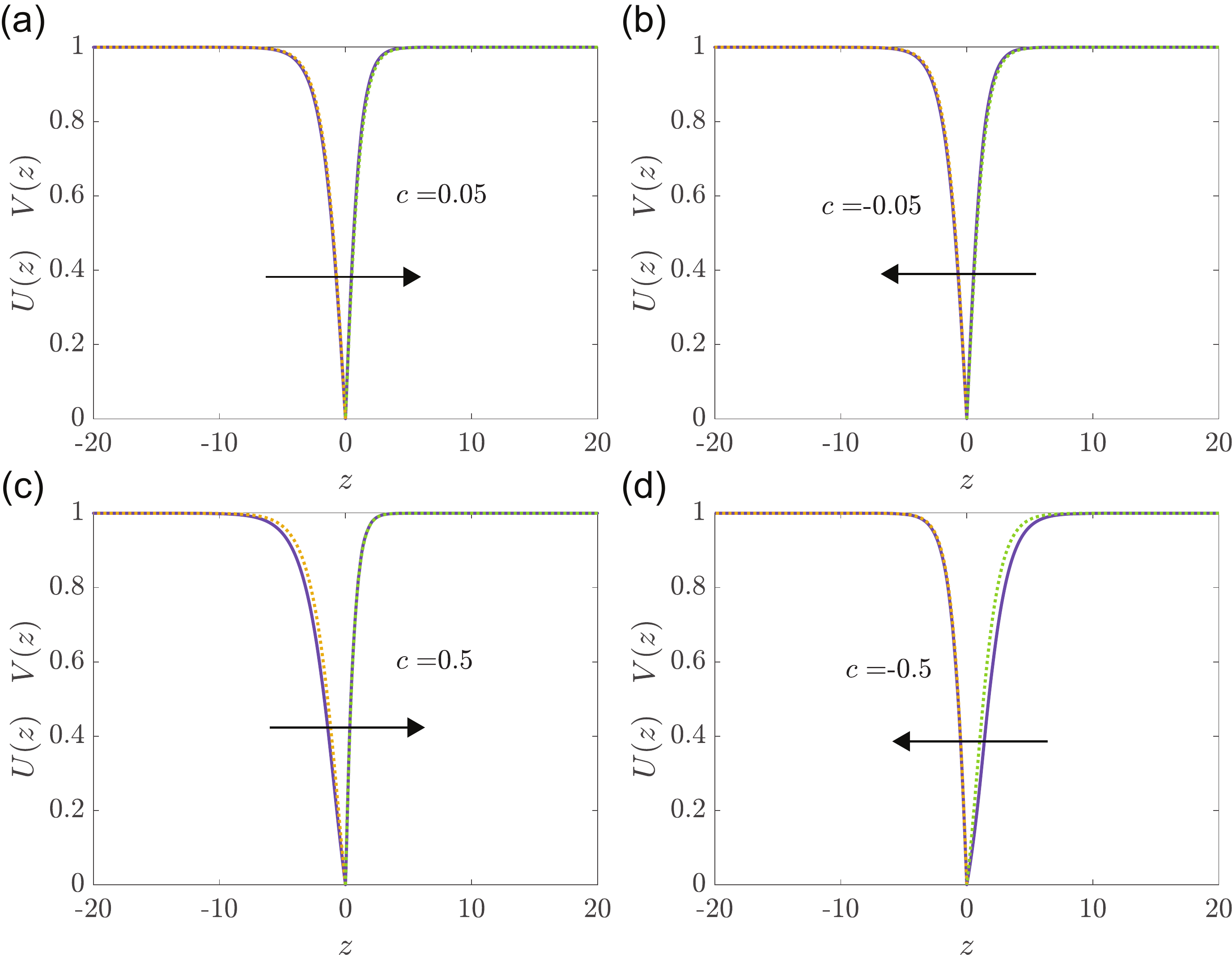}
	\caption{\textbf{Perturbation solution for the shape of the travelling waves when $D = 0.5$ and $\lambda = 1$.} Comparison of $U(z)$ and $V(z)$ from the $\mathcal{O}(c)$ perturbation solution (purple solid) with numerical estimates obtained by solving Equations (16)-(17) that are shifted so that $U(0)=V(0)=0$.  Numerical estimates of $U(z)$ and $V(z)$ are shown in dashed yellow and dashed green lines, respectively.  Results are shown for: (a) $c=0.05$; (b) $c=-0.05$; (c) $c=0.5$; and (d) $c=-0.5$.}
	\label{fig:figS10}
\end{figure}

\begin{figure}[H]
	\centering
	\includegraphics[width=1\textwidth]{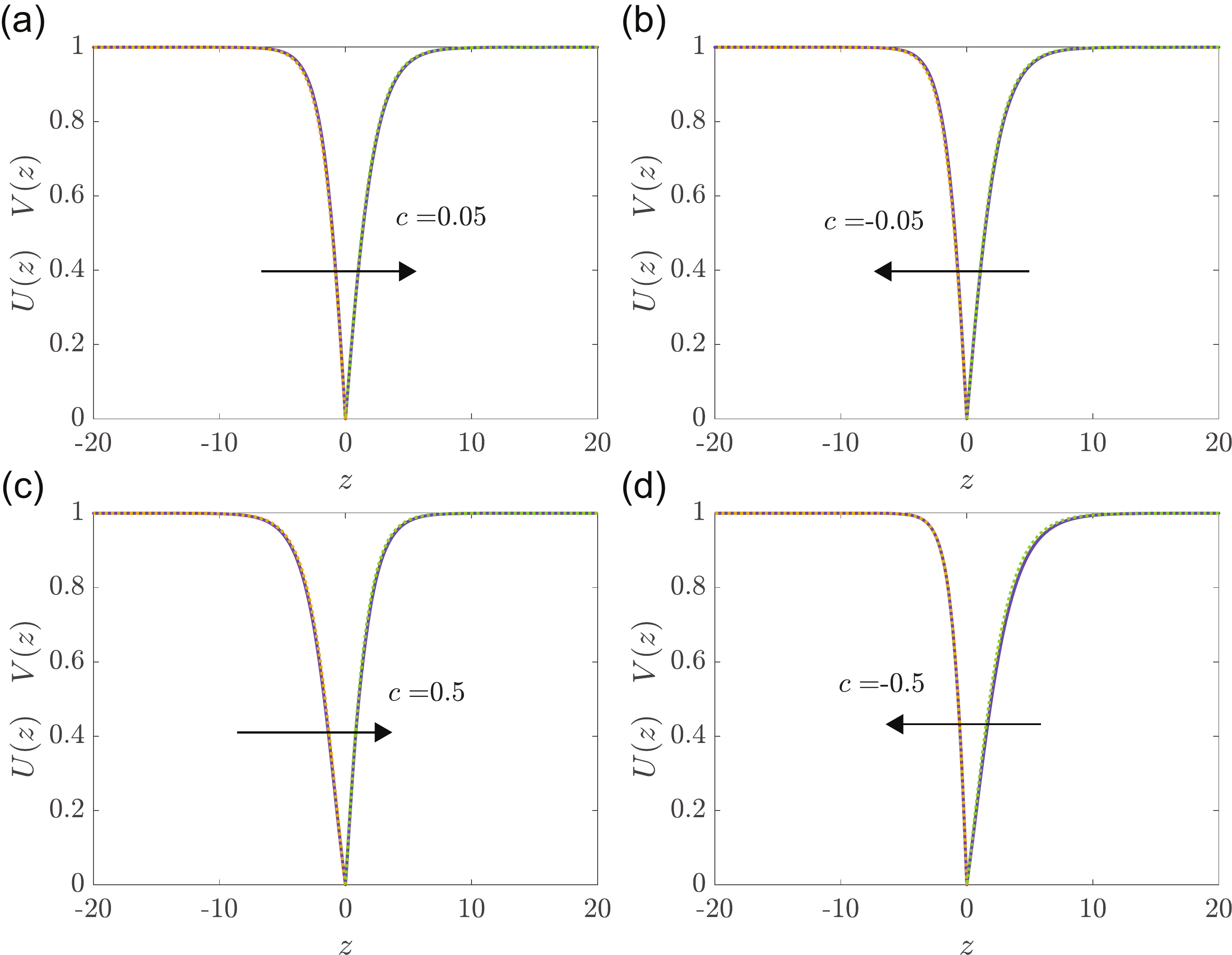}
	\caption{\textbf{Perturbation solution for the shape of the travelling waves when $D = 2$ and $\lambda = 1$.} Comparison of $U(z)$ and $V(z)$ from the $\mathcal{O}(c)$ perturbation solution (purple solid) with numerical estimates obtained by solving Equations (16)-(17) that are shifted so that $U(0)=V(0)=0$.  Numerical estimates of $U(z)$ and $V(z)$ are shown in dashed yellow and dashed green lines, respectively.  Results are shown for: (a) $c=0.05$; (b) $c=-0.05$; (c) $c=0.5$; and (d) $c=-0.5$.}
	\label{fig:figS11}
\end{figure}

\begin{figure}[H]
	\centering
	\includegraphics[width=1\textwidth]{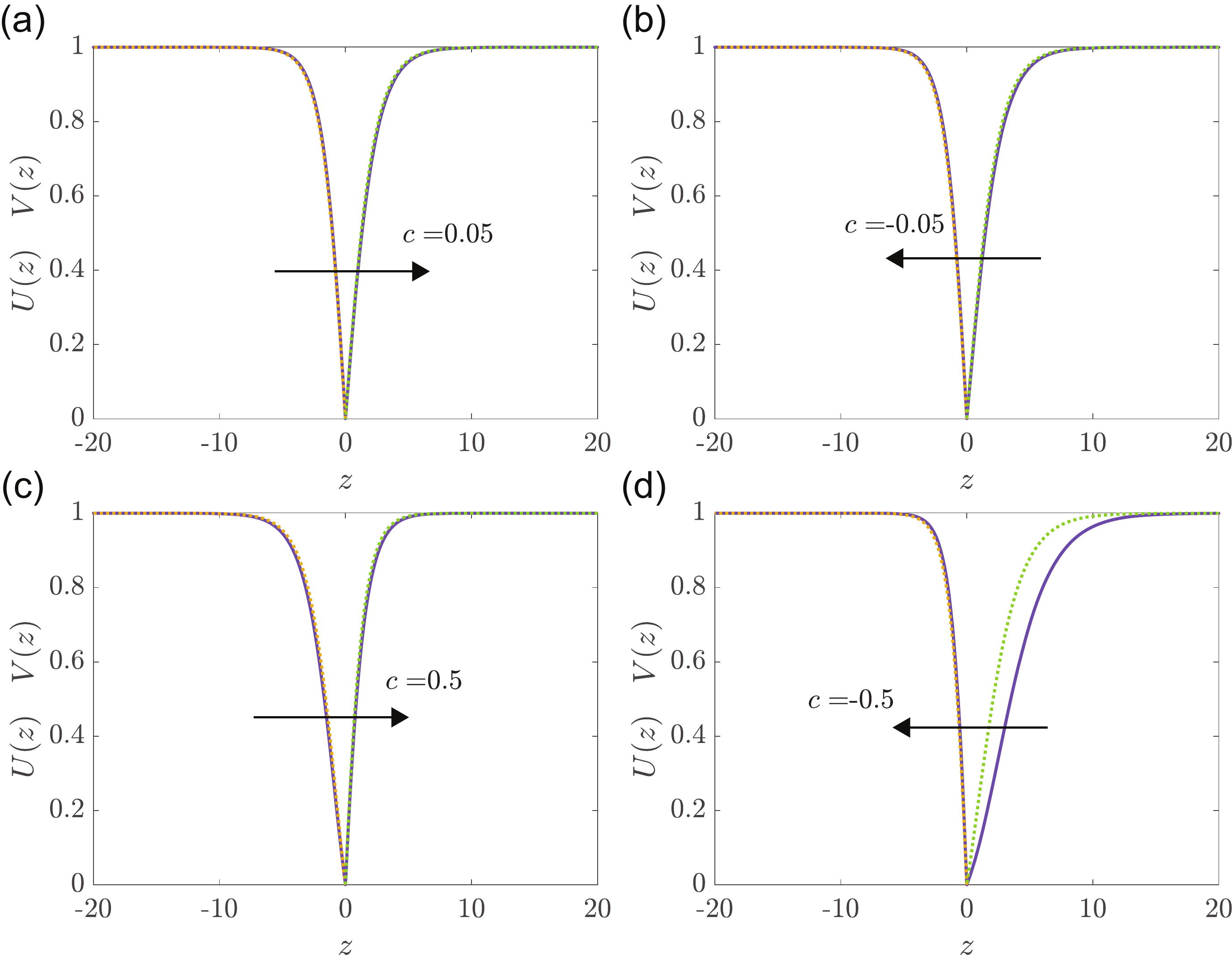}
	\caption{\textbf{Perturbation solution for the shape of the travelling waves when $D = 1$ and $\lambda = 0.5$.} Comparison of $U(z)$ and $V(z)$ from the $\mathcal{O}(c)$ perturbation solution (purple solid) with numerical estimates obtained by solving Equations (16)-(17) that are shifted so that $U(0)=V(0)=0$.  Numerical estimates of $U(z)$ and $V(z)$ are shown in dashed yellow and dashed green lines, respectively.  Results are shown for: (a) $c=0.05$; (b) $c=-0.05$; (c) $c=0.5$; and (d) $c=-0.5$.}
	\label{fig:figS12}
\end{figure}

\begin{figure}[H]
	\centering
	\includegraphics[width=1\textwidth]{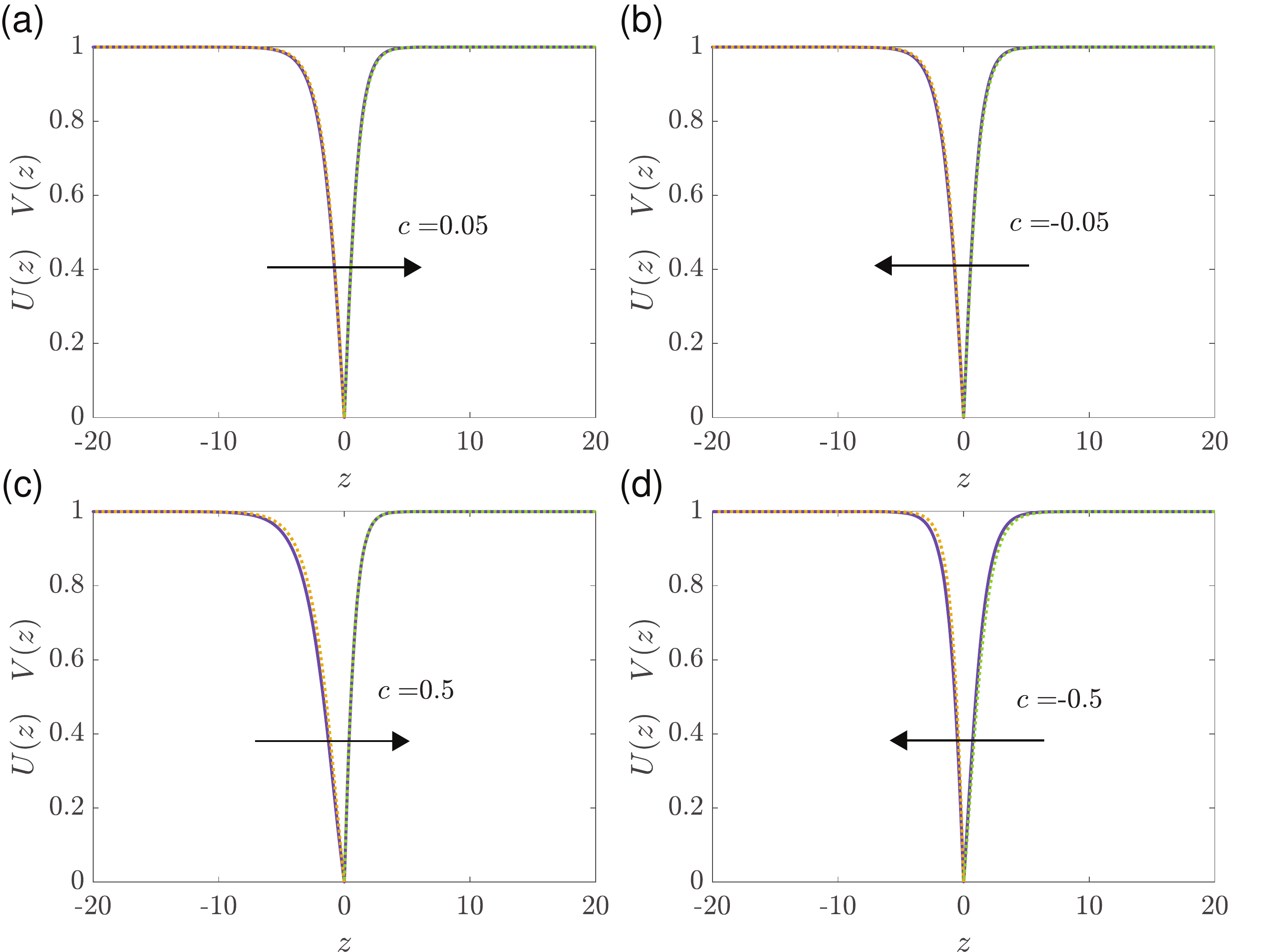}
	\caption{\textbf{Perturbation solution for the shape of the travelling waves when $D = 1$ and $\lambda = 2$.} Comparison of $U(z)$ and $V(z)$ from the $\mathcal{O}(c)$ perturbation solution (purple solid) with numerical estimates obtained by solving Equations (16)-(17) that are shifted so that $U(0)=V(0)=0$.  Numerical estimates of $U(z)$ and $V(z)$ are shown in dashed yellow and dashed green lines, respectively.  Results are shown for: (a) $c=0.05$; (b) $c=-0.05$; (c) $c=0.5$; and (d) $c=-0.5$.}
	\label{fig:figS13}
\end{figure}

\end{document}